\documentclass[twocolumn,prd, superscriptaddress, showpacs]{revtex4-1}
\usepackage{hyperref,xcolor}
\usepackage{amsmath,amssymb}
\usepackage{graphicx}
\usepackage{bm}

\newcommand{\Z}{\ensuremath{\mathbb{Z}}}

\newcommand{\rd}{\ensuremath{\mathrm{d}}}

\providecommand{\abs}[1]{\lvert#1\rvert}

\newcommand{\expv}[1]{\left\langle#1\right\rangle}

\newcommand{\bk}{\ensuremath{\bm{k}}}
\newcommand{\bphi}{\bm{\Phi}}
\newcommand{\bdphi}{\bm{\delta\Phi}}
\newcommand{\be}{\begin{equation}}
\newcommand{\ee}{\end{equation}}
\newcommand{\benn}{\nonumber\begin{equation}}
\newcommand{\eenn}{\nonumber\end{equation}}
\def\bea{\begin{eqnarray}} \def\eea{\end{eqnarray}}
\def\beann{\begin{eqnarray*}} \def\eeann{\end{eqnarray*}}
\def\lsim{\raise0.3ex\hbox{$<$\kern-0.75em\raise-1.1ex\hbox{$\sim$}}}
\def\gsim{\raise0.3ex\hbox{$>$\kern-0.75em\raise-1.1ex\hbox{$\sim$}}}

\begin{document}

\title{\vspace*{-1.5cm}
{\hfill \texttt{\footnotesize CERN-PH-TH/2013-097}}
\vfill
Dynamical Mean Field Approximation Applied to Quantum Field Theory}
\date{\today}
\author{Oscar Akerlund}
\affiliation{Institut f\"ur Theoretische Physik, ETH Zurich, CH-8093 Zurich, Switzerland}
\author{Philippe de Forcrand}
\affiliation{Institut f\"ur Theoretische Physik, ETH Zurich, CH-8093 Zurich, Switzerland}
\affiliation{CERN, Physics Department, TH Unit, CH-1211 Geneva 23, Switzerland}
\author{Antoine Georges}
\affiliation{Centre de Physique Th\'eorique, CNRS, \'Ecole Polytechnique, 91128 Palaiseau, France}
\affiliation{Coll\`ege de France, 11 place Marcelin Berthelot, 75005 Paris, France}
\affiliation{DPMC, Universit\'e de Gen\`eve, 24 quai Ernest Ansermet, CH-1211 Geneva, Switzerland}
\author{Philipp Werner}
\affiliation{Department of Physics, University of Fribourg, 1700 Fribourg, Switzerland}

\begin{abstract}
We apply the Dynamical Mean Field (DMFT) approximation to the real, scalar $\varphi^4$ quantum field theory. By comparing to lattice Monte Carlo calculations, perturbation theory and standard mean field theory, we test the quality of the approximation in two, three, four and five dimensions. The quantities considered in these tests are the critical coupling for the transition to the ordered phase and the associated critical exponents $\nu$ and $\beta$. We also map out the phase diagram in the most relevant case of four dimensions. In two and three dimensions, DMFT incorrectly predicts a first order phase transition for all bare quartic couplings, which is problematic, because the second order nature of the phase transition of lattice $\varphi^4$-theory is crucial for taking the continuum limit. Nevertheless, by extrapolating the behavior away from the phase transition, one can obtain critical couplings and critical exponents. They differ from those of mean field theory and are much closer to the correct values. 
In four dimensions the transition is second order for small quartic couplings and turns weakly first order as the coupling increases beyond a tricritical value. In dimensions five and higher, DMFT gives qualitatively correct results, predicts reasonable values for the critical exponents and considerably more accurate critical couplings than standard mean field theory. The approximation works best for small values of the quartic coupling. We investigate the change from first to second order transition in the local limit of DMFT which is computationally much cheaper. We also discuss technical issues related to the convergence of the non-linear self-consistency equation solver and the solution of the effective single-site model using Fourier-space Monte Carlo updates in the presence of a $\varphi^4$-interaction.
\end{abstract}

\maketitle

\section{Introduction}\label{sec:introduction}\noindent
The numerical simulation of quantum field theories, typically in $d=4$ 
dimensions (3-space + Euclidean time), is a major computational challenge. 
Simulations suffer severely from the high cost of increasing the
4-dimensional lattice volume. 
Moreover, in interesting situations like real-time evolution or non-zero
matter density, a ``sign problem'' arises, which makes the computational cost
grow exponentially with the volume.
In such cases, we are forced to consider simplified models which, at best, capture only the most relevant properties of the full model. One such approximation is to consider a Mean Field version of the theory in question. 
The simplest mean field approach reduces the problem to a zero-dimensional one where the field, or the gauge-invariant plaquette in gauge theories, is allowed to fluctuate in the background of a self-consistently determined mean field that represents the influence of the field at all other points in space-time. Mean field theories have been an important tool in the study of field theories for a long time. From the Ising model to QCD \cite{Batrouni:1982bg, Batrouni:1982dx, Batrouni:1984rc, Kallman:1984ky, Green:1983sd},  mean field approximations give us hints about phase transitions and critical behavior. Although there exist regions in parameter space where mean field theory gives very good or even exact results (usually when $d=\infty$), it is obviously a very crude approximation in most regions of physical parameters. Hence, it is desirable to go beyond mean field theory and to develop an approximation which provides a better description of fluctuations. 

An approach which has proven very useful for the study of correlated lattice models relevant for solid state physics is \emph{Dynamical Mean Field Theory (DMFT)} \cite{Georges:1996zz, Georges:lec_notes_dmft}. Here, the word ``dynamical" refers to the fact that the mean field can fluctuate in one direction, typically the Euclidean time direction, while remaining constant in the $(d-1)$ other dimensions. The $d$-dimensional lattice problem is thus mapped onto a one-dimensional problem with non-local interactions representing the influence of the remaining degrees of freedom.
Due to this dynamical dimension, DMFT gives access to correlation functions which are localized in the frozen directions. If the dynamical dimension is the imaginary-time axis, DMFT furthermore enables the calculation of finite-temperature expectation values. Obtaining access to this kind of information is an additional motivation to explore the DMFT approach.

In general, the effective one-dimensional model must be solved numerically, for example using a  (Quantum) Monte Carlo method \cite{Gull:2011}. As in mean field theory, the DMFT calculation involves a self-consistent computation of the (dynamical) mean field, which in practice amounts to solving a set of non-linear equations self-consistently. The increased complexity arises from the fact that the field to be optimized is a function (or a collection of functions) of one variable. 

DMFT was initially developed for fermionic systems, but the theory has recently been extended and successfully applied to bosonic lattice systems \cite{Byczuk:2008, Anders:2010, Anders:2011} and bose-fermi mixtures \cite{Anders:2012}. The bosonic version of DMFT can, with rather straightforward modifications, be applied to the $\varphi^4$ quantum field theory. It is thus an interesting question how well this approach, which manifestly breaks Lorentz invariance, can capture the phase diagram and critical behavior of lattice field theories.

In models where local interactions dominate it is a reasonable further approximation to study the local limit of DMFT.  
The effective model then reduces to a single site problem with two coupled self-consistency equations. 
This provides a generalization of standard mean-field theory, in which both the first and second moment of the field are 
self-consistently determined \cite{Pankov:2002}. 

The structure of the paper is as follows. We briefly introduce $\varphi^4$ theory in section~\ref{sec:phi4}. We then discuss the mean field approximation and DMFT in sections~\ref{sec:mf}~and~\ref{sec:dmft}. In section~\ref{sec:mc} we discuss the Monte Carlo method that we use to solve the effective single-site model. In section \ref{sec:emft} we briefly discuss the local limit of DMFT. Section~\ref{sec:res} presents the numerical results for  $\varphi^4$-theory in dimensions 2 to 5. We give a short summary and an outlook on possible extensions in section~\ref{sec:sum}.

\section{$\varphi^4$ Theory}\label{sec:phi4}\noindent
$\varphi^4$ theories are an important class of quantum field theories. Even the simplest incarnation, with a real scalar field, exhibits interesting phenomena like spontaneous symmetry breaking (SSB) with a second order phase transition. One important application is in the Standard Model Higgs sector, which consists of a two-component complex $\varphi^4$ theory, but the interest in such theories extends far beyond that. Because of their relative simplicity, $\varphi^4$ theories are often used as a testing ground and stepping stone when developing new methods. We will explicitly discuss here the real scalar $\varphi^4$ theory, but the approach can readily be generalized to complex fields (see Appendix A).

The Lagrangian density of real scalar $\varphi^4$ theory reads
\be
\mathcal{L}[\varphi(x)] = \frac{1}{2}\partial_\mu \varphi(x)\partial^\mu \varphi(x) - \frac{1}{2}m_0^2\varphi(x)^2 - \frac{g_0}{4!}\varphi(x)^4,
\ee
using a $d$-dimensional Minkowski metric, $(+,-,\ldots,-)$. This model is a prototype for spontaneous symmetry breaking: here, the $\Z_2$ global symmetry, $\varphi(x) \leftrightarrow -\varphi(x) ~\forall x$, is spontaneously broken for negative [renormalized] $m^2$ via a second order phase transition at $m^2=0$. After Wick rotating time to the imaginary axis to obtain a Euclidean metric, we discretize the action and apply the conventional change of variables:
\bea
a^{\tfrac{d-2}{2}}\varphi(x) &=& \sqrt{2\kappa}\varphi_x,\\
(am_0)^2 &=& \frac{1-2\lambda}{\kappa} - 2d,\label{eq:mgkl_m}\\
a^{4-d}g_0 &=& \frac{6\lambda}{\kappa^2}.\label{eq:mgkl_g}
\eea
The action can then be defined on a regular $d$-dimensional hypercubic lattice with an extent $L=aN_l$ in each direction, so that there are $N=N_l^d$ lattice sites. The action expressed in terms of $\kappa$ and $\lambda$ becomes
\be\label{eq:action_latt}
S = \sum_x\left(-2\kappa\sum_\mu\varphi_{x+\widehat{\mu}}\varphi_x + \varphi_x^2 + \lambda(\varphi_x^2-1)^2\right).
\ee
The renormalized mass and coupling are unknown functions of $\kappa$ and $\lambda$, which must be determined via numerical simulations or perturbation theory. Since we can only measure dimensionless observables, the renormalized, physical mass $m_R$ always appears together with a factor of the lattice spacing $a$. Keeping the physical mass fixed, this implies that a second order phase transition on the lattice, i.e. $(am_R)\to0$, in fact \emph{defines} the continuum limit $a\to0$. Thus, our interest in the lattice model is focused on the behavior in the vicinity of the phase transition corresponding to the spontaneous $\Z_2$ symmetry breaking, when one approaches the transition both from the symmetric and from the broken-symmetry phase.

\section{Mean Field Theory}\label{sec:mf}\noindent
The Mean Field approximation has been an important tool in the study of field theories for a long time. The idea behind mean field theory is to simplify the model by mapping it to a zero-dimensional effective model, which means that all interactions except contact terms are replaced by an interaction with a constant background field. For $\varphi^4$ theory on the lattice, the partition function is
\be
Z \!= \!\int\!\mathcal{D}[\varphi]\prod_x\exp\left(\!-\varphi_x^2 \!-\! \lambda(\varphi_x^2-1)^2 \!+\!2\kappa \sum_{\mu=1}^d\varphi_x\varphi_{x+\widehat{\mu}}\!\right).
\ee
By fixing the field to $m$ at all lattice sites but one, we find the self-consistency equation 
\begin{align}
m \equiv \expv{\varphi} =& \frac{1}{Z_{MF}}\int_{-\infty}^{\infty}\rd\varphi\,\varphi\exp\Big(-\varphi^2 - \lambda(\varphi^2-1)^2\nonumber\\
& \hspace{3.25cm}+2\kappa(2d)m \varphi\Big),\label{eq:MF}\\
Z_{\rm MF} =& \int_{-\infty}^{\infty}\rd\varphi\,\exp\left(-\varphi^2 - \lambda(\varphi^2-1)^2 +2\kappa(2d)m \varphi\right).
\end{align}
The critical coupling, $\kappa_c$, can be expressed in terms of modified Bessel functions after an expansion of $\exp(2\kappa(2d)m\varphi)$ for small values of $m$, but the expression is not very enlightening. Using the same expansion it is also easy to check that $m\propto(\kappa-\kappa_c)^{1/2}$, which implies the critical exponent $\beta=1/2$.

In the limit $\lambda\to +\infty$, we have $\varphi(x)=\pm 1 \equiv \sigma(x)$ and the lattice field-theory reduces to 
an Ising model:
\begin{equation} 
Z^\text{Ising} =\sum_{\{\sigma\}}\exp\left(2\kappa\sum_{\expv{i,j}}\sigma_i\sigma_j \right).
\end{equation}
Standard mean field theory maps the $d$-dimensional problem onto a zero-dimensional model coupled to a constant (Weiss) effective field: 
\begin{equation}
Z^\text{Ising}_{\rm MF} = \sum_{\sigma = \pm1}\exp{(\beta h_{\rm eff}\sigma)}\,=\,2\cosh \beta h_{\rm eff}.
\end{equation}
The effective field, $h_\text{eff}$, is fixed by the self-consistency equation,
\be
\expv{\sigma} = \frac{1}{2d}h_\text{eff} \Rightarrow  \expv{\sigma} = \tanh\left[2d(2\kappa) \expv{\sigma}\right].
\ee

Beside DMFT, A number of methods have been developed that systematically improve on mean field theory. One example is the extended mean field theory, which we will introduce as the local limit of DMFT in Sec.~\ref{sec:emft}. Other noteworthy examples are the so-called ``Cluster variation methods'' \cite{Bethe:1935, Kikuchi:1951, Morita:1994}. These methods are systematic extensions of mean field theory, which in the limit of large clusters approach the exact result, but they break translation invariance and do not give direct access to correlation functions. We do not pursue these methods further here.

\section{Dynamical Mean Field Theory}\label{sec:dmft}\noindent
We will now introduce DMFT as an extension of standard mean field theory. There are many parallels between the two approximations as the name suggests, but there are also differences. Our derivation of the one-dimensional effective model in the disordered and broken-symmetry phases is analogous to that for bosonic DMFT (see Refs.~\cite{Anders:2010} and \cite{Anders:2011}). DMFT extends on mean field theory by treating an effective field which is a function of one variable, i.e. the full dynamics is preserved in one direction, while fluctuations in the $(d-1)$ others are frozen. Since we are working with a Lorentz invariant field theory it makes no difference which direction is singled out, but we will follow the convention from solid state physics and call the dynamical dimension $t$ with conjugate momentum $\omega$, and the other dimensions $x_1,\ldots, x_{d-1}$ with conjugate momenta $k_1,\ldots, k_{d-1}$.
Moreover, in this way, the finite-temperature behavior of our quantum field 
theory can be studied by varying the extent of the dynamical dimension.

\begin{figure}[bh!]
\centering
\includegraphics[width=1\linewidth]{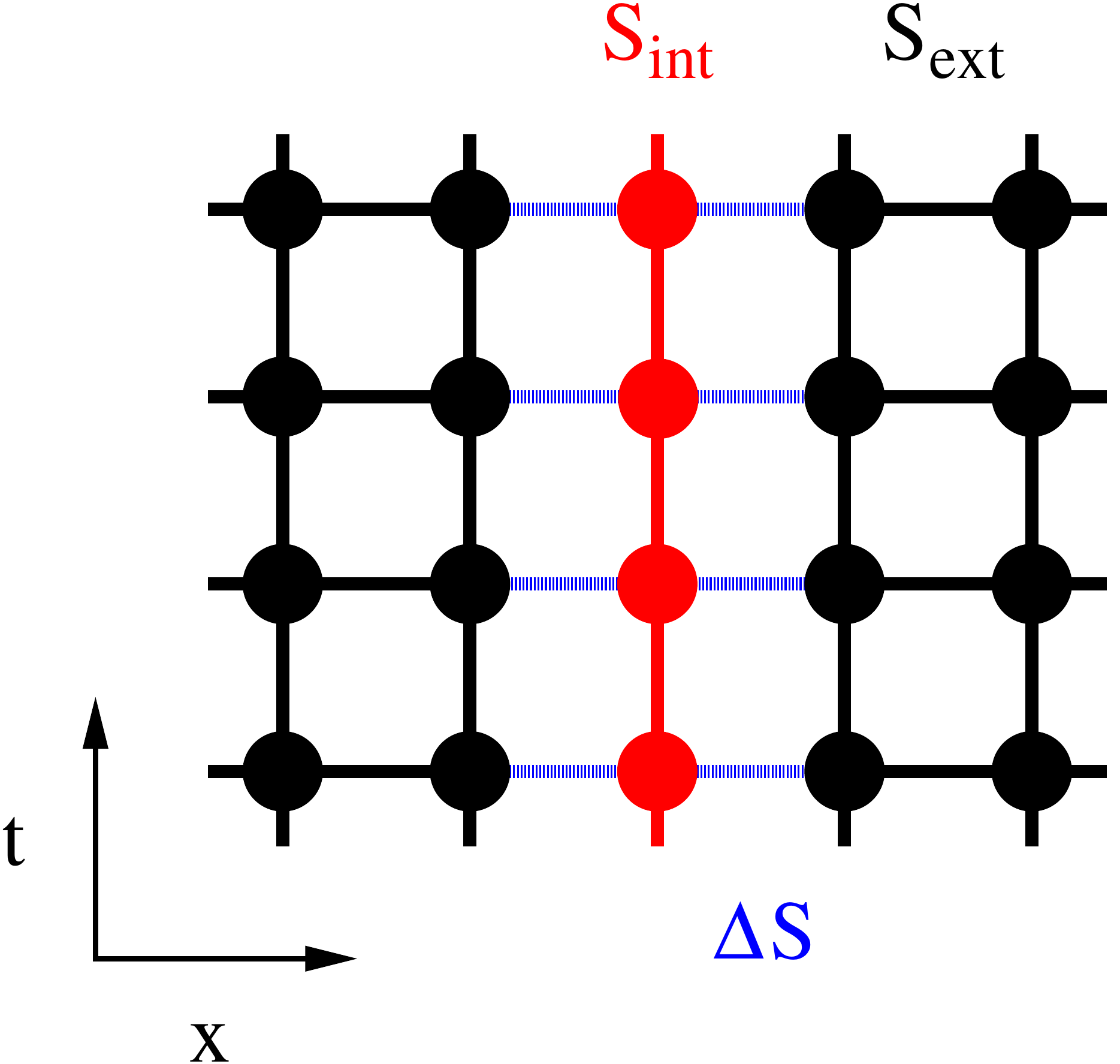}
\caption{Graphical interpretation of the cavity method where the action \eqref{eq:delta_S} is decomposed into an internal part, $S_\text{int}$, an external part, $S_\text{ext}$, and an interaction part, $\Delta S$. The external degrees of freedom are then integrated out after a Taylor-expansion of the interaction, $\exp(-\Delta S)$.}
  \label{fig:cavity}
\end{figure}

Since DMFT takes full account of the fluctuations along the time direction, the effective model is defined on a one-dimensional chain, with non-local couplings in time. Schematically, starting from a model with purely local potential $V$,
\be
Z = \int \mathcal{D}[\varphi]\exp\left(\sum_x\sum_{\mu}\varphi_x\varphi_{x+\widehat{\mu}} - \sum_xV(\varphi_x)\right),
\ee
one obtains
\begin{align}
Z_{\text{DMFT}}
&= \int \mathcal{D}[\varphi]\exp\Bigg(-\sum_{t,t'}\varphi_tK^{-1}(t-t')\varphi_{t'} \nonumber\\
&\hspace{2.6cm} \!-\! \sum_tV(\varphi_t) + h\sum_t\varphi_t\Bigg),\label{eq:Z_dmft}
\end{align}
where $K$ is a non-local effective kernel which emulates the propagation in the $(d-1)$ frozen dimensions. We will here briefly discuss how the effective one-dimensional model is obtained. In Appendix~\ref{app:imp} we present a complete derivation of the effective action for complex $\varphi^4$ theory. The idea is to split the degrees of freedom into {\em internal} ($\varphi_{\text{int},t} \equiv \varphi_{(\vec{0},t)}$) and {\em external} ($\varphi_{\text{ext},t} \equiv \{\varphi_{(\vec{y},t)}; \vec{y} \neq \vec{0}\}$) degrees of freedom, and integrate out the latter 
(cavity method) \cite{Georges:1996zz, Anders:2011}. The action 
(\ref{eq:action_latt})
separates into three parts: $S = S_\text{int} + \Delta S + S_\text{ext}$, with
\bea
S_\text{int} &=&  \displaystyle\sum_t\Big[-2\kappa\varphi_{\text{int},t +1}\varphi_{\text{int},t} + \varphi_{\text{int},t}^2 + \lambda(\varphi_{\text{int},t}^2-1)^2\Big],\nonumber\\
\Delta S &=& -2\kappa \displaystyle\sum_t \varphi_{\text{int},t}\sum_{\langle \text{int},\text{ext}\rangle} \varphi_{\text{ext},t}, \label{eq:delta_S}\\
S_\text{ext} &=&  \displaystyle\sum_{x\neq(\vec{0},t)}\Big[-2\kappa\sum_{\substack{\nu\\x+\hat\nu\neq(\vec{0},t)}} \varphi_{x +\widehat{\nu}}\varphi_x + \varphi_x^2 + \lambda(\varphi_x^2-1)^2\Big].\nonumber
\eea
In the second equation, the sum over $\langle\text{int,ext}\rangle$ is shorthand for the sum over all external sites at time $t$ which are nearest neighbors to the internal site at time $t$. See Fig.~\ref{fig:cavity} for a graphical interpretation of the decomposition.

The next step is to Taylor-expand $\exp\left(-\Delta S\right)$ and integrate out the external degrees of freedom $\varphi_{\text{ext},t}$. This will give a series of expectation values of $n$-point functions of external fields, \linebreak$\expv{\varphi_{\text{ext},t_1}\cdots\varphi_{\text{ext},t_n}}_\text{ext}$, which can be re-exponentiated to produce effective couplings among $n$ internal fields. Depending on how high the powers of connected correlators we keep, we obtain more or less complicated self-consistency equations. It is an easy exercise to check that keeping only $\expv{\varphi_{\text{ext},t}}_\text{ext}$ yields standard mean field theory, i.e. a linear coupling to a constant background (due to translational invariance). In DMFT we normally keep both the linear terms, which introduce a coupling to an effective external field,
\be\label{eq:phi_ext}
\expv{\varphi_{\text{ext},t}}_\text{ext}\varphi_{\text{int},t} \Rightarrow \phi_\text{ext}\varphi_t,
\ee
and the quadratic fluctuations, which give rise to a non-local quadratic term
\be\label{eq:delta_t}
\varphi_{\text{int},t}\expv{\varphi_{\text{ext},t}\varphi_{\text{ext},t'}}_{\text{ext}}\varphi_{\text{int},t'} \Rightarrow \varphi_t\Delta(t-t')\varphi_{t'}.
\ee
In general, also higher order fluctuations could be considered, which would introduce interactions among three, four, etc. fields. These non-local interactions are due to propagation through the effective medium. In the limit where all orders are taken into account, the mapping to the one-dimensional model becomes exact. Each term kept in the expansion of $\exp(-\Delta S)$ introduces one (non-local) coupling that has to be determined self-consistently. From Eqs.~\eqref{eq:phi_ext} and \eqref{eq:delta_t} we find that the following quantities are related:
\begin{align}
\phi_\text{ext} &\leftrightarrow \expv{\varphi}_{Z_\text{DMFT}}, \\
\Delta(t-t') &\leftrightarrow \expv{\varphi_t\varphi_{t'}}_{Z_\text{DMFT}}.
\end{align}
In fact, the correspondence between $\phi_\text{ext}$ and $ \expv{\varphi}_{Z_\text{DMFT}}$ must be an equality, whereas the connection between $\Delta(t-t')$ and the two point correlator is more involved and will be discussed below.

In anticipation of a broken symmetry it is more convenient to expand the fields around their expectation value when deriving the effective action (see Appendix~\ref{app:imp}). This implies that we will work only with connected quantities which will be labeled with a subscript $c$. In our case we consider up to quadratic fluctuations and the resulting one-dimensional model 
will from now on be referred to as the ``impurity model", and quantities related to it will be subscripted with an `imp'. 
This terminology follows the established DMFT terminology in the context of condensed-matter physics: the full lattice model is 
mapped onto a lower-dimensional entity (`impurity') coupled to a self-consistent environment. Here, the `impurity' is a one-dimensional world-line of a single spatial site:    
\begin{align}
&S_\text{imp} = \sum_{t,t'}\varphi_t K_{\text{imp},c}^{-1}(t-t')\varphi_{t'} + \lambda\displaystyle\sum_t (\varphi_{t}^2-1)^2 - h\sum_t\varphi_t,\label{eq:simp_1}\\
&\widetilde{K}_{\text{imp},c}^{-1}(\omega) = 1-2\kappa\cos(\omega) - \widetilde{\Delta}(\omega),\label{eq:simp_2}\\
&h = 2\phi_\text{ext}(2\kappa(d-1)-\widetilde{\Delta}(0)).\label{eq:simp_3}
\end{align}
$\widetilde{K}_{\text{imp},c}^{-1}(\omega)$ is the inverse of the connected two-point Green's function of the free ($\lambda=0$) theory and $h$ is an effective external magnetic field, which is non-zero in the broken-symmetry phase (See Appendix \ref{app:imp}). (We put a tilde on Fourier transformed quantities.)

The frequency-dependent effective coupling $\widetilde{\Delta}(\omega)$ is determined self-consistently by demanding that 
the impurity Green's function coincides with the local propagator of the full model. Quite generally we can express the Green's function of some interacting theory in momentum space as
\be\label{eq:dyson_latt}
\widetilde{G}(\bk,\omega) = \frac{1}{\widetilde{G}_0^{-1}(\bk,\omega) + \widetilde{\Sigma}(\bk,\omega)},
\ee
where $\widetilde{G}_0^{-1}(\bk,\omega) = 1-2\kappa\sum_{i=1}^d\cos(k_i)$ is the Green's function of the free $d$-dimensional theory and $\widetilde{\Sigma}$ is the self-energy which captures the interaction effects. The ``local" Green's function, from $\vec{x}=\vec{0}$ to $\vec{x}=\vec{0}$, 
is obtained by summing over all spatial momenta,
\be\label{eq:dyson_latt_loc}
\widetilde{G}_\text{loc}(\omega) = \sum_{\bk}\frac{1}{\widetilde{G}_0^{-1}(\bk,\omega) + \widetilde{\Sigma}(\bk,\omega)},
\ee
where the momentum sum is normalized such that $\sum_{\bk} 1 =1$. The Green's function of the impurity model also satisfies such a relation,
\be\label{eq:dyson_imp}
\widetilde{G}_\text{imp}(\omega) = \frac{1}{\widetilde{K}_{\text{imp},c}^{-1}(\omega) + \widetilde{\Sigma}_\text{imp}(\omega)}.
\ee  
DMFT approximates the exact self-energy $\widetilde{\Sigma}$ with the self-energy $\widetilde{\Sigma}_\text{imp}$ of the impurity system, i.e. $\widetilde{\Sigma}(\bk,\omega)\approx\widetilde{\Sigma}_\text{imp}(\omega) = \widetilde{G}_\text{imp}^{-1}(\omega)-\widetilde{K}_{\text{imp},c}^{-1}(\omega)$, which can be substituted in Eq.~\eqref{eq:dyson_latt_loc}. The local Green's function may then be expressed as 
\be\label{eq:dyson_dmft}
\widetilde{G}_\text{loc}(\omega) = \sum_{\bk}\frac{1}{\widetilde{G}_0^{-1}(\bk,\omega) + \widetilde{G}_\text{imp}^{-1}(\omega) - \widetilde{K}_{\text{imp},c}^{-1}(\omega)},\hspace{4mm}
\ee
or, alternatively, in terms of $\widetilde{\Delta}(\omega)$, as
\be
\widetilde{G}_\text{loc}(\omega) = 
\sum_{\bk}\frac{1}{\widetilde{G}_\text{imp}^{-1}(\omega)+\widetilde{\Delta}(\omega) 
- 2\kappa\sum_{i=1}^{d-1}\cos k_i}.\label{eq:dyson_dmft_delta}
\ee

\begin{figure}[t]
\centering
\includegraphics[width=1\linewidth]{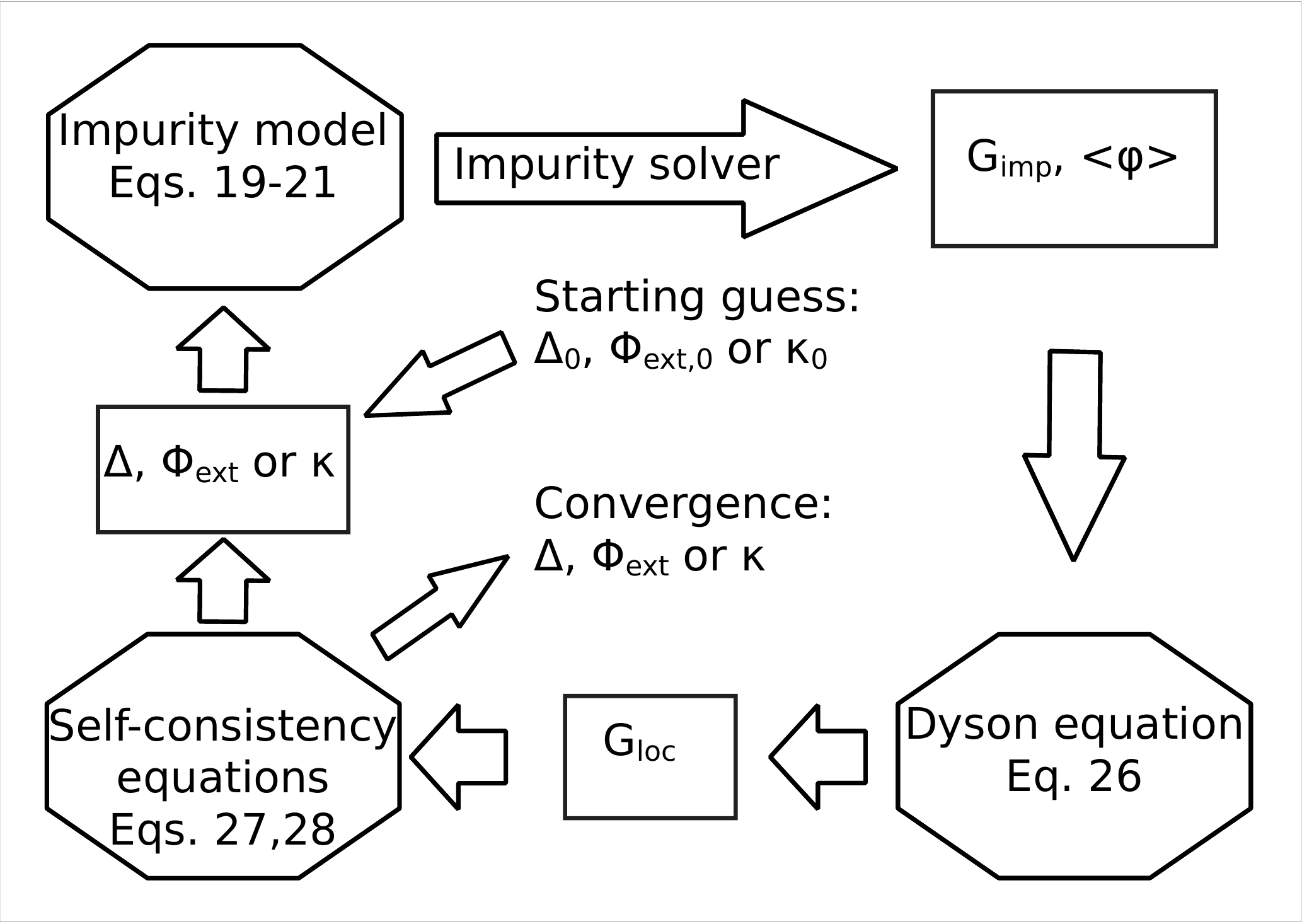}
\caption{Schematic depiction of the DMFT procedure. For a given quartic coupling $\lambda$ and either $\kappa$ or $\phi_\text{ext}$ fixed we make a guess for $\Delta$ and the non-fixed variable. This defines an impurity action via Eqs.~(\ref{eq:simp_1}-\ref{eq:simp_3}). We then solve this effective model for the Green's function and the expectation value of the field, $\expv{\varphi}$. The local Green's function of the full model is approximated via Eq.~\eqref{eq:dyson_dmft_delta}. The self-consistency equations, (\ref{eq:sc_delt},\ref{eq:sc_mean}), are then used to calculate new values for $\Delta$ and $\phi_\text{ext}$ or $\kappa$. This procedure is repeated until the self-consistency equations are satisfied.}
  \label{fig:dmft_flow}
\end{figure}

The self-consistency condition identifies the local Green's function \eqref{eq:dyson_latt_loc} with the impurity Green's function \eqref{eq:dyson_imp}, which thus implicitly determines $\widetilde{K}_{\text{imp},c}(\omega)$ (or $\widetilde{\Delta}$). The two coupled self-consistency equations then read
\begin{align}
\widetilde{G}_\text{imp}(\omega) &= \widetilde{G}_\text{loc}(\omega),\label{eq:sc_delt}\\
\expv{\varphi}_{S_\text{imp}} &= \phi_\text{ext}.\label{eq:sc_mean}
\end{align}
The DMFT procedure is illustrated as a circular flowchart in Fig.~\ref{fig:dmft_flow}.

\section{Impurity Solver}\label{sec:mc}
\noindent
As shown in the flowchart of Fig.~\ref{fig:dmft_flow}, a DMFT calculation requires the repeated evaluation of $G_\text{imp}$ and $\expv{\varphi}$ for successive values of the effective interaction $K_{\text{imp},c}$, so that both the accuracy and the efficiency of the ``impurity solver" are relevant issues. We use a Monte Carlo method, which allows to reach arbitrary precision in polynomial time. 

The general form of the impurity action is
\be
S_\text{imp} = \displaystyle\sum_{t,t'} \varphi_t K_{\text{imp},c}^{-1}(t-t')\varphi_{t'} - h\sum_t\varphi_t + \sum_tV(\varphi_t),\nonumber
\ee
where $K^{-1}_{\text{imp},c}$ is non-local but translation invariant, and thus diagonal in
momentum space. $V(\varphi_t)$ is local, and thus diagonal in position space, and goes to $+\infty$ as $\varphi$ goes to $\pm\infty$. The non-local nature of the kernel renders single-site updates inefficient, while the potential $V$ prevents a local formulation in Fourier space. To overcome this difficulty, we use the method proposed in Ref.~\cite{Matsuo:2006}. It substitutes the exponential of the potential at each time-slice by a sum of $M$ different Gaussians,
\be\label{eq:mc_jap}
\exp[-V(\varphi_t)] \approx \sum_{m=1}^M\mu_m\exp\left[-\nu(\varphi_t-\sigma_m)^2\right].
\ee
It is important to keep the width, $\nu$, independent of $m$, so that the quadratic part of the action (for a given selection of Gaussians, one per $t$-value) will be translation invariant, and thus diagonal in Fourier space. The number of Gaussians, $M$, is chosen empirically and the other parameters are determined via fitting for fixed $M$. As $M\to\infty$, the sum over $m$ turns into an integral and the Gaussians turn into delta functions, so that Eq.~\eqref{eq:mc_jap} becomes an equality. It is therefore desirable to use a large $M$ to keep the error in the approximation small but at the same time keep $M\ll N_t$ to make the updates more efficient than single-site updates. 

The variables $\mu_m$ and $\sigma_m$ play the role of auxiliary variables and the partition function can be written as
\be
Z = \sum_{\{m_t\}}\int\mathcal{D}[\varphi]W(\{\varphi\},\{m_t\}).
\ee
The update is then performed in two steps. First, for fixed $\varphi$, each time-slice $t$ is assigned a Gaussian term $m_t$ according to the heat-bath probability,
\be
p(m_t|\varphi_t) = \frac{ \mu_{m_t}\exp[-\nu(\varphi_t-\sigma_{m_t})^2]}{\sum_{m}\mu_{m}\exp[-\nu(\varphi_t-\sigma_{m})^2]}.
\ee
Then, for fixed $\{m_t\}$, $\varphi$ is updated. The point of this update scheme is that for fixed $\{m_t\}$ the action is quadratic in $\varphi$ and translation invariant, i.e. diagonal in Fourier space. We have $W(\{\varphi\},\{m_t\}) = \exp(-\hat{S})$ with
\begin{align}
\hat{S} =& \frac{1}{N_t}\sum_n\left[\widetilde{K}_{\text{imp},c}^{-1}(\omega_n)\abs{\tilde{\varphi}_n}^2 + \nu\abs{\tilde{\varphi}_n-\tilde{\sigma}_n}^2\right] -h\tilde{\varphi}_0\nonumber\\
=&  \frac{1}{N_t}\sum_n\left(\nu+\widetilde{K}_{\text{imp},c}^{-1}(\omega_n)\right) \left|\tilde{\varphi}_n-\frac{\nu\tilde{\sigma}_n+\tfrac{N_t}{2}h\delta_{n,0}}{\nu+\widetilde{K}_{\text{imp},c}^{-1}(\omega_n)}\right|^2,
\end{align}
up to terms that do not depend on $\varphi$. A new configuration can now efficiently be generated by sampling the Gaussian distribution which is defined by $\hat{S}$. The total complexity of updating all $N_t$ components of $\varphi$ is $N_t\text{max}\left(\log(N_t),M\log(M)\right)$, $N_t\log N_t$ from Fourier transforming $\sigma$ and $\varphi$ and $N_tM\log M$ from searching the list of heat bath probabilities, $p(m_t|\varphi_t)$ for all $t$. 
In cases where the effective field $h$ is zero or small it can be advantageous to combine this update scheme with cluster updates to sample the configuration space more efficiently. There exist appropriate cluster methods that can deal with non-local interactions and ``double-well" potentials, see Ref.~\cite{Werner:2005}. However, if the external field is large (as is the case for example if the symmetry breaking transition is first order), the cluster updates become inefficient. 

\section{Solutions of the self-consistency equations}\label{sec:sols}
\begin{figure}[t]
\centering
\includegraphics[width=1\linewidth]{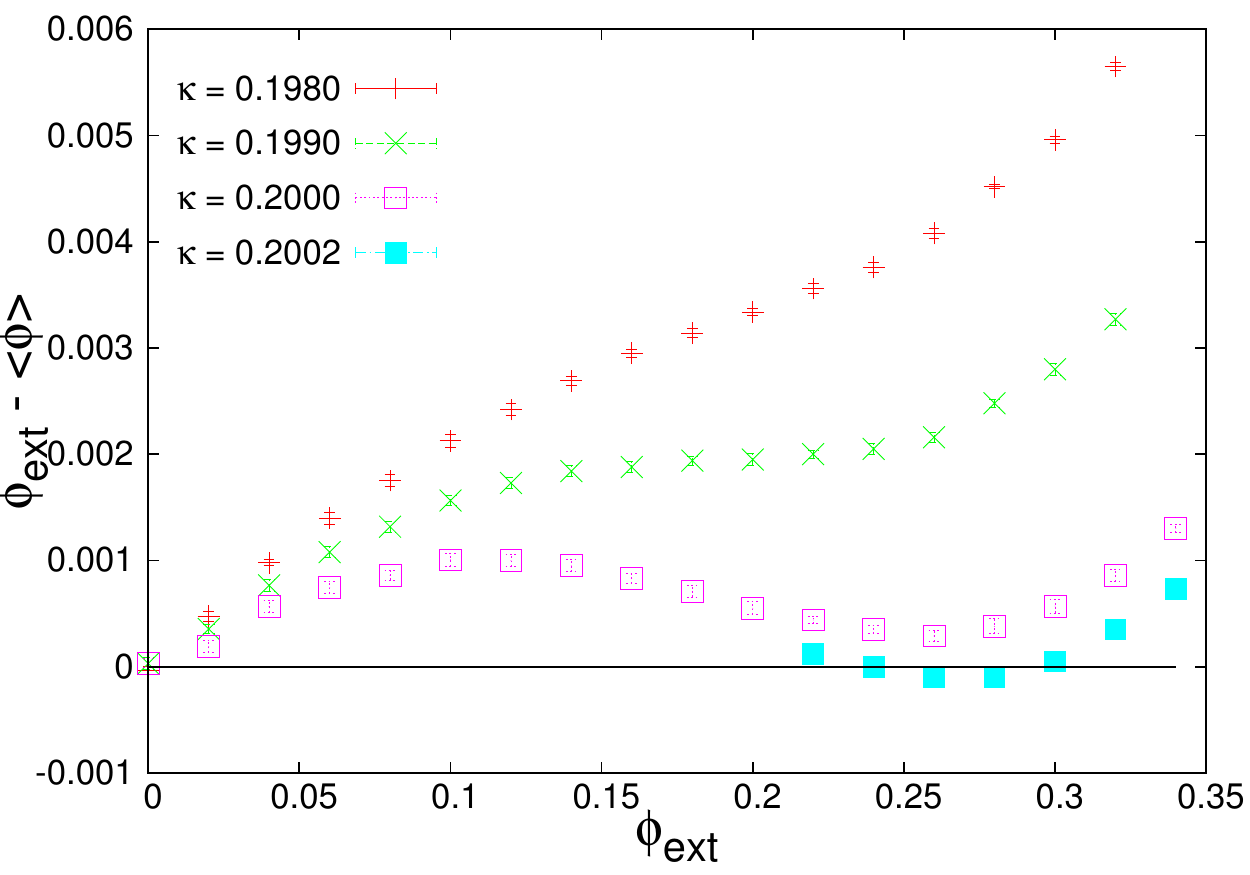}
\caption{Results for $d=3$ and $\lambda=0.5$, showing the change in $\phi_{\text{ext}}-\expv{\varphi}$ in an iterative substitution scheme when solving Eq.~\eqref{eq:dyson_dmft_delta}. For each data point we fix $\phi_{\text{ext}}$ and iterate $\widetilde{\Delta}$ to convergence and then measure $\expv{\varphi}$. A line going through zero with a positive (negative) slope is a stable (unstable) fixed point. We see clearly the first order transition at $\kappa\approx0.2002$. It is also evident that the change in $\phi_{\text{ext}}$ might be extremely slow just above this coupling which might incorrectly be interpreted as convergence.}
  \label{fig:dphi_3d}
\end{figure}

\noindent
By studying the solution of the self-consistency equations very close to the phase transition we have found that for dimensions lower than five DMFT wrongly gives a first order transition, while Monte Carlo simulations of the full model correctly give a second order transition for all dimensions $d\geq 2$. The first order behavior can be hard to detect for weak quartic couplings. Conventional iterative substitution methods to solve the self-consistency equations are also not well suited to detect such behavior, which may thus be overlooked. This is because forward-substitution methods can only find stable fixed-points and slow convergence together with statistical noise can conceal a small first-order jump. We propose an alternative update procedure which speeds up convergence and allows us to obtain all fixed points of Eq.~\eqref{eq:dyson_latt_loc}, even unstable ones. An illustrative example of the first order behavior in three dimensions can be seen in Fig.~\ref{fig:dphi_3d}. There, we plot the deviation of $\phi_{\text{ext}}$ from the self-consistent solution, 
by iteratively solving Eq.~\eqref{eq:dyson_latt_loc} in three dimensions for fixed $\phi_{\text{ext}}$. Stable (unstable) fixed points  
correspond to zero crossings 
with positive (negative) slope. This figure clearly demonstrates the first order transition, and also a potential problem arising from a possibly very slowly converging $\phi_{\text{ext}}$.

Our solution to this problem is to interchange the roles of $\kappa$ and $\phi_{\text{ext}}$. Instead of fixing $\kappa$ and searching for stable fixed points, we fix $\phi_{\text{ext}}$ and search for the unique $\kappa$ giving a self-consistent solution of Eqs.~(\ref{eq:sc_delt},\ref{eq:sc_mean}), i.e. we search for the root of
\be
f\left(\left.\widetilde{\Delta}(\omega_n),\kappa\right|\phi_{\text{ext}}\right) = \left(\widetilde{G}_{\text{loc}}(\omega_n)-\widetilde{G}_{\text{imp}}(\omega_n),\phi_{\text{ext}}-\expv{\varphi}\right).
\ee
Fast convergence can be achieved by using generalized Newton methods with either an approximated or numerically exact Jacobian matrix. In our Monte Carlo scheme it is straightforward and cheap to directly sample the Jacobian. This approach to solving the self-consistency equations has much in common with the phase space-extension used by Strand et al. \cite{Strand:2011} to study the first order Mott transition in the Hubbard model.

\section{Extended Mean Field Theory: a local limit of DMFT}
\label{sec:emft}
\begin{figure}[t]
\centering
\includegraphics[width=1\linewidth]{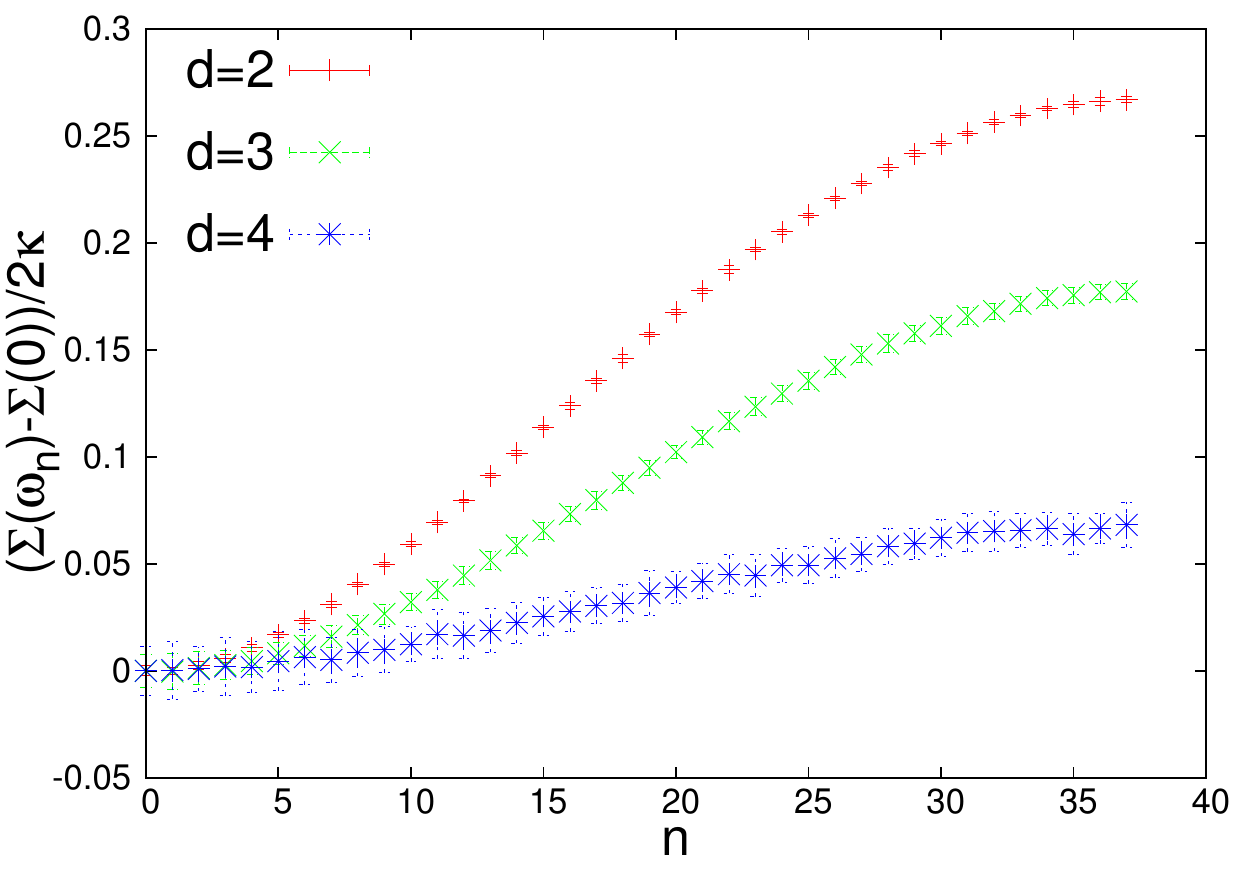}
\caption{The frequency-dependent part of the self-energy, normalized to that of
the inverse free propagator (i.e. divided by $2\kappa$)), for $\lambda = 1$ 
($d=2,3$) and $\lambda=2$ ($d=4$).
The self-energy is only weakly frequency-dependent, especially for larger
dimension, and can be further approximated by a constant, in the EMFT approximation.
}
  \label{fig:self_energy}
\end{figure}

\noindent
At a second order phase transition the full lattice Green's function, Eq.~\eqref{eq:dyson_latt}, becomes massless, i.e $\widetilde{G}(0,\omega)\propto \omega^{-2},\; \omega\to 0$.
This does not, however, imply that also $\widetilde{G}_\text{loc}$ (Eq.~\eqref{eq:dyson_dmft_delta}) and $\widetilde{G}_\text{imp}$ (Eq.~\eqref{eq:dyson_imp}) behave similarly. In fact we find that the phase transition is mainly driven by a large contact term in $\Delta(t-t')$, which cancels the mass in $G_\text{imp}$, rather than a long-range tail which could trigger spontaneous symmetry breaking \cite{Dyson:1969up, Dyson:1969up_2, Frohlich:1982gf, Fisher:1972zz, Luijten:2001}.
We further observe that the self-energy 
$\widetilde{\Sigma}_\text{imp}(\omega) = \widetilde{\Sigma}(\omega)$
also shows only a mild dependence on $\omega$, especially in higher dimensions,
as illustrated in Fig.~\ref{fig:self_energy}.
These two observations motivate us to simplify DMFT further and consider its local version in which 
$\widetilde{\Delta}$ and $\widetilde{\Sigma}$ are frequency-independent,
and thus local in $t$: $\Delta(t-t') = \Delta \delta_{t,t'}$.
The DMFT construction then reduces to the scheme introduced 
by Pankov, Kotliar and Motome \cite{Pankov:2002} in their study of so-called extended DMFT.
In this local limit, the impurity action simplifies to  
\be
S_\text{imp}= (1-\Delta)\varphi^2 -2\phi_\text{ext}(2d\kappa-\Delta)\varphi + \lambda(\varphi^2-1)^2.
\ee
It involves two variational parameters, $\Delta$ and $\phi_\text{ext}$, while standard mean-field theory involves only the 
order parameter $\phi_\text{ext}$. 
The Green's function $G_\text{imp}$ is now just a number, which is the variance of the field: 
\be
G_\text{imp} = 2(\expv{\varphi^2}-\expv{\varphi}^2).
\ee
$G_\text{imp}$ is purely local in space and time and is required to coincide with the full Green's function $G(\bm{r},t)$ at the origin $(\bm{0},0)$. The expression for the local Green's function Eq.~\eqref{eq:dyson_dmft} stays the same but with an additional integral over $\omega$:
\begin{align}
G_\text{loc} &= \int\frac{\rd k^d}{(2\pi)^d}\left[G_\text{imp}^{-1}+\Delta - 2\kappa\sum_{i=1}^d \cos(k_i)\right]^{-1}\nonumber\\
&= \int_0^\infty \rd \tau \exp\left[-\tau\left(G_\text{imp}^{-1} + \Delta\right)\right]I_0(2\kappa\tau)^d, \label{eq:G_EMFT}
\end{align}
where $I_0(x)$ is the zeroth modified Bessel function. Demanding self-consistency, i.e. $\expv{\varphi} = \phi_\text{ext}$ and $G_\text{imp} = G_\text{loc}$, leads to a set of two coupled integral equations that can easily be solved numerically and compared to the DMFT result. This self-consistent scheme will from here on be referred to as \emph{Extended Mean Field Theory} or EMFT.

Note the difference with the single self-consistency equation Eq.~\eqref{eq:MF} of
the standard mean-field treatment: here, mass renormalization is made possible
via the parameter $\Delta$, which is coupled with the wave-function renormalization
via Eq.~\eqref{eq:G_EMFT}. As we will see, this improved but still local approximation
provides a dramatic improvement in the estimate of the critical coupling $\kappa_c$.

\begin{figure}
\centering
\includegraphics[width=1\linewidth]{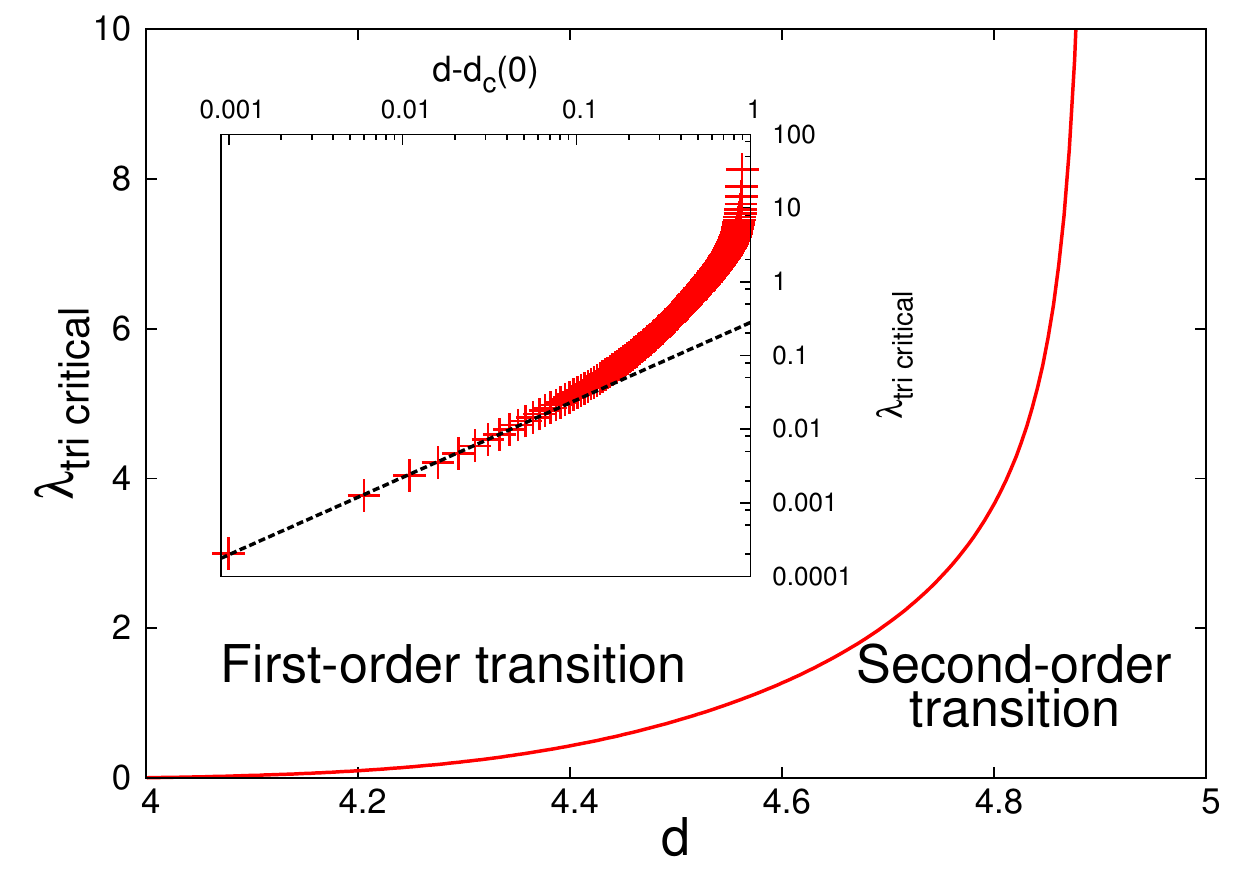}
\caption{The tricritical quartic coupling for EMFT as a function of dimension. The critical dimension at $\lambda=0$ is $d_c(0) \approx 4.00$. The inset shows a power law behavior as $d$ approaches $d_c(0)$.
}
  \label{fig:lambda_tc}
\end{figure}

Depending on the dimension and the value of the quartic coupling, we get either a first or second order transition. The tricritical coupling, $\lambda_{\text{tc}}$ where the order of the transition changes is shown in Fig.~\ref{fig:lambda_tc}. For $d=4$ the tricritical $\lambda$ is found to be $0$. The critical $\kappa$ is found by solving the following equation for $\kappa$,

\begin{align}
\expv{\varphi^2}_{S_c} &= \frac{I_d}{4\kappa},\\
S_c &= \left(1-\frac{2\kappa}{I_d}\left(dI_d -1\right)\right)\varphi^2 - \lambda(\varphi^2-1)^2,\\
I_d &\equiv \int\rd \tau e^{-\tau d}I_0(\tau)^d,
\end{align}
where $I_0(x)$ is the zeroth modified Bessel function of the first kind. In the Ising limit this simplifies to 
\be
\kappa_c(d) = \frac{I_d}{4}.
\ee
A detailed treatment can be found in Appendix~\ref{app:sdmft}.

\section{Results}\label{sec:res}\noindent
We judge the quality of the DMFT approximation by how well it reproduces the critical coupling, $\kappa_c$, and the critical exponents $\beta$ and $\nu$ compared to Monte Carlo and standard mean field theory. In the case of a first order transition we define the critical coupling as if the transition was second order, i.e. we fit the magnetization to a power law. In the DMFT loop we measure the field expectation value, $\expv{\varphi}$, and the renormalized mass, $(a m_R)$. The latter is extracted from the Green's function in momentum space,
 \be
Z_R(m_R^2+\omega^2) \approx \widetilde{G}_c^{-1}(\bm{0},\omega) = \widetilde{G}_{c,\text{imp}}^{-1}(\omega) + \widetilde{\Delta}(\omega) - 2(d-1)\kappa,
\ee
for a range of $\omega$ close to zero. This procedure is more robust against noise than the usual second moment definition of the mass where only the data points in $\omega=0$ and $\omega=2\pi/N_t$ are considered. $Z_R^{1/2}$ is the wave function renormalization. From $m_R$ and $\expv{\varphi}$ we can extract $\kappa_c$, $\beta$ and $\nu$ via the fits
\bea
\xi = \frac{1}{m_R} &\propto& (\kappa-\kappa_c)^{-\nu}, \label{eq:mass_nu}\\
\expv{\varphi} &\propto& (\kappa-\kappa_c)^{\beta}.\label{eq:phi_beta}
\eea
We will present results successively in $5$, $4$, $3$ and $2$ dimensions to show how the quality of the approximation depends on the dimensionality. 

\begin{figure}[t]
\centering
\includegraphics[width=1\linewidth]{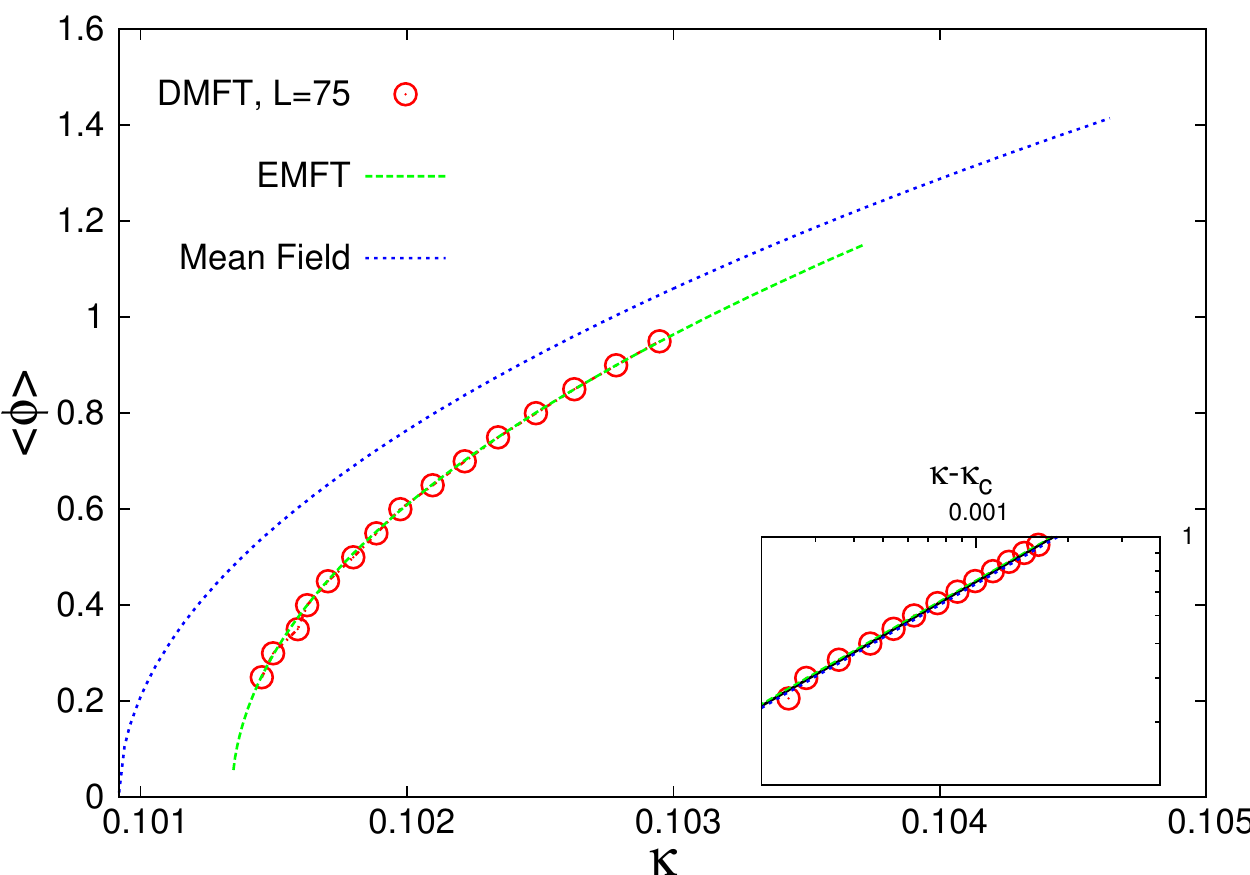}\\
\includegraphics[width=1\linewidth]{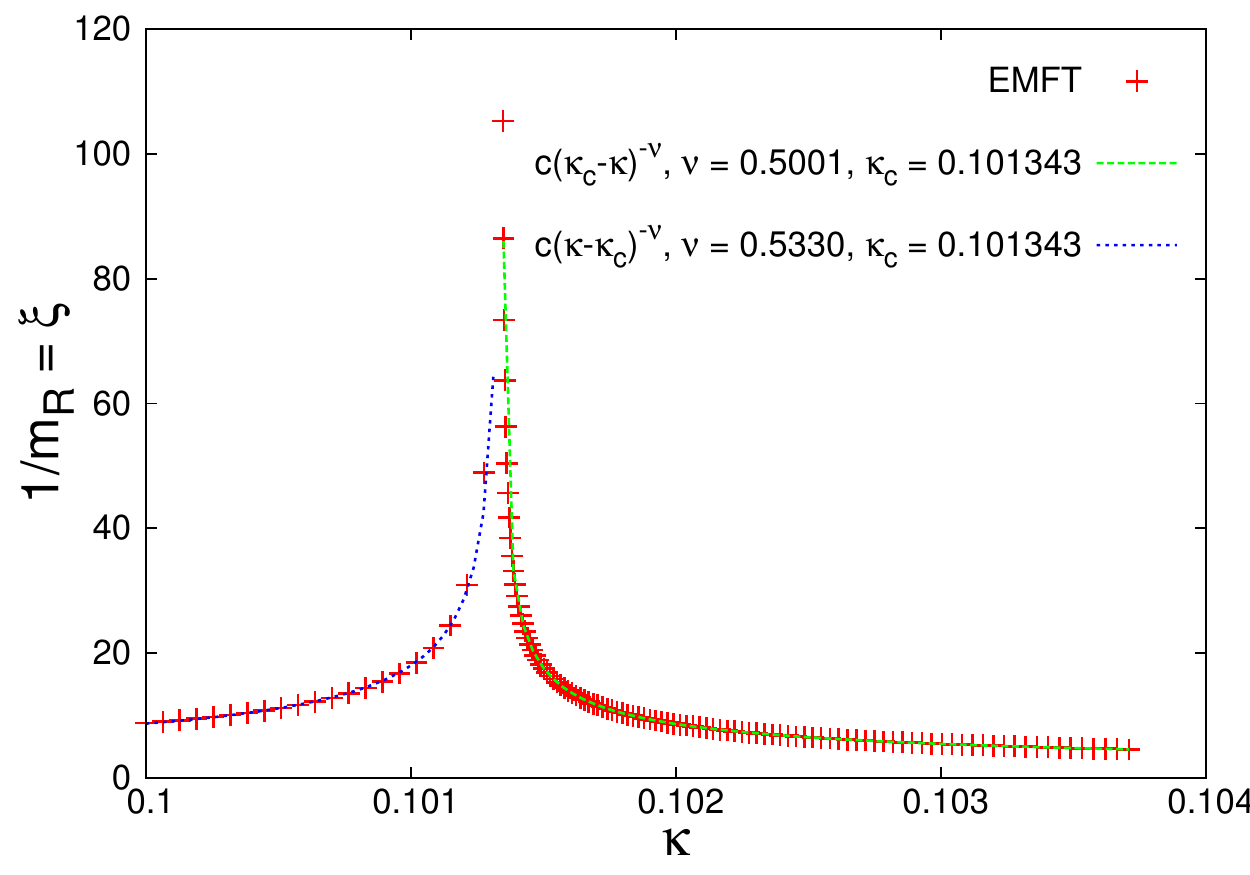}
\caption{Results for $d=5$ and $\lambda = 0.01$.
\emph{Top panel}: Expectation value of $\varphi$ as a function of the coupling, $\kappa$, obtained by DMFT, EMFT and mean field theory. The inset shows a log-log plot in the broken-symmetry phase. The slope of the line gives the critical exponent $\beta$. The best fit to Eq.~\eqref{eq:phi_beta} is given by $\beta=0.499(4)$ and $\kappa_c = 0.101345(5)$ for DMFT and $\beta=0.500(1)$ and $\kappa_c = 0.1013425(3)$ for EMFT.
\emph{Bottom panel}: Correlation length, $\xi=1/m_R$, obtained from EMFT as a function of the coupling, $\kappa$. In the broken-symmetry phase we find $\nu = 0.500(1)$ with fixed $\kappa_c=0.1013425$ and in the symmetric phase we find $\nu = 0.533(4).$}
  \label{fig:expv_5d_l001}
\end{figure}

\noindent
$\bullet$ In five dimensions we have only standard mean field results to compare with, so we can only guess if DMFT improves the value of the critical coupling. It is however known that mean field theory always underestimates the critical coupling $\kappa_c$ and we find that DMFT gives a larger $\kappa_c$ than mean field theory. In Fig.~\ref{fig:expv_5d_l001} we show that DMFT predicts critical exponents, $\beta=0.499(4)$ and $\nu = 0.504(4)$, which are very close to the exact values $\beta=\nu=1/2$. The correlation length obtained by DMFT agrees with the one obtained by EMFT. The values of $\beta$, $\nu$ and $\kappa_c$ have been obtained by minimizing the Chi-square of a linear fit of log-log data to Eqs.~\eqref{eq:mass_nu} and \eqref{eq:phi_beta}. We also see that there is quite a remarkable agreement between DMFT and EMFT. Moreover, the simpler EMFT is not affected by numerical errors as DMFT close to the transition.

\begin{figure}[t]
\centering
\includegraphics[width=1\linewidth]{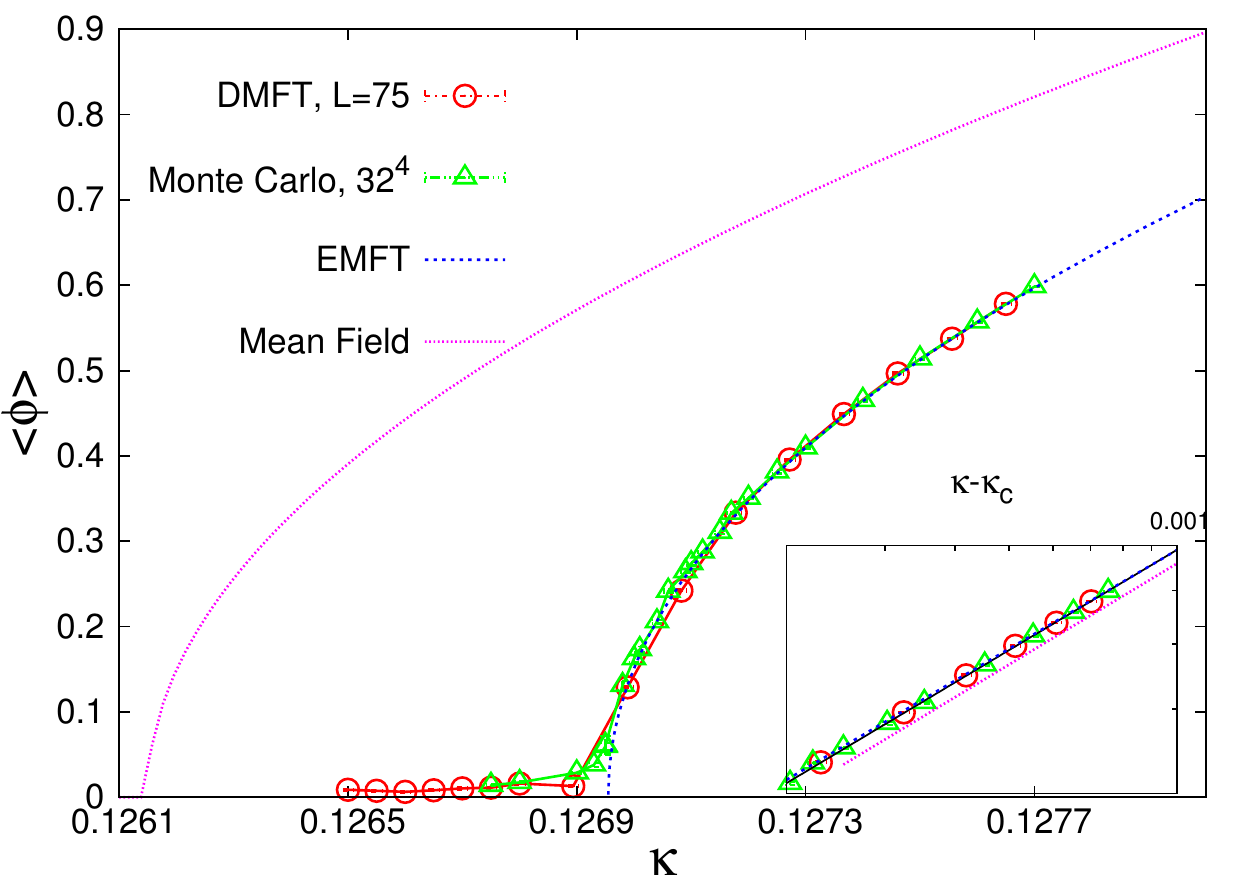}\\
\includegraphics[width=1\linewidth]{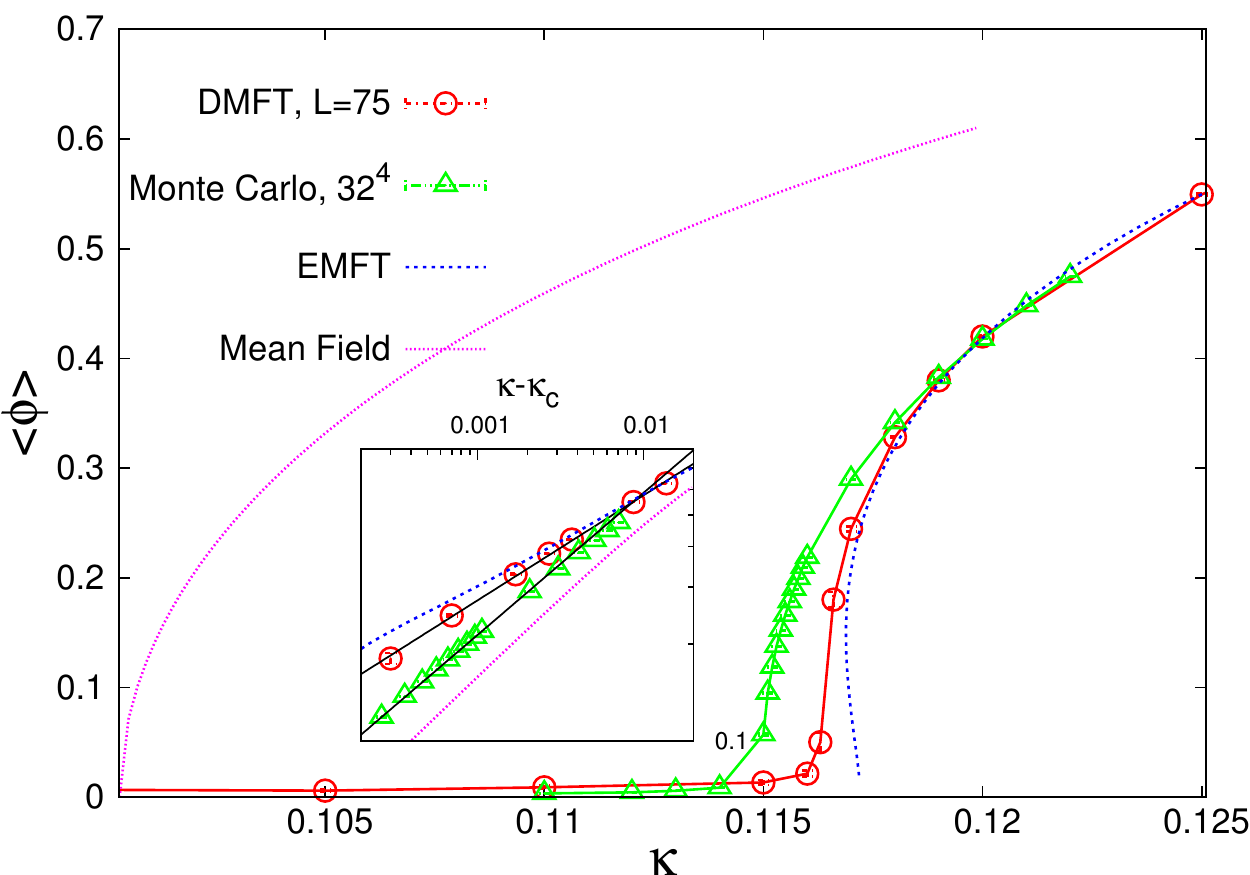}
\caption{Results for $d=4$, $\lambda=0.01$ (\emph{top panel}) and $\lambda=2$ (\emph{bottom panel}). Expectation value of $\varphi$ as a function of the coupling, $\kappa$, obtained by DMFT, Monte Carlo simulation and mean field theory. The inset shows a log-log plot of the broken-symmetry phase. From the slope of the fitted line we extract the critical exponent $\beta$.  
}
  \label{fig:expv_4d_l001}
\end{figure}

\noindent
$\bullet$ Next, we consider the four-dimensional theory. Here, we can compare with Monte Carlo simulations of the full theory, where we used a mixture of local updates and cluster updates of Wolff-type \cite{Wolff:1989}. Each data point is typically obtained from $10^5$ measurements, where $2N_l^d$ local updates and $20$ cluster updates were performed between successive measurements (the integrated auto-correlation time was $\tau \approx 10$ measurements). In Fig.~\ref{fig:expv_4d_l001}
we show the expectation value of the field for weak ($\lambda=0.01$) and strong ($\lambda = 2$) coupling. 
For $\lambda=0.01$, DMFT gives $\beta=0.508(5)$ and $\kappa_c=0.1269475(5)$, while the Monte Carlo data yield $\beta=0.497(3)$ and $\kappa_c=0.1269470(4)$. In the strong coupling case, $\lambda=2$, the best fit of the DMFT data to Eq.~\eqref{eq:phi_beta} gives $\beta=0.325(4)$ and $\kappa_c=0.1163(3)$, whereas from the Monte Carlo data, one obtains $\beta=0.44(7)$ and $\kappa_c=0.1144(5)$.

\begin{figure}
\includegraphics[width=1\linewidth]{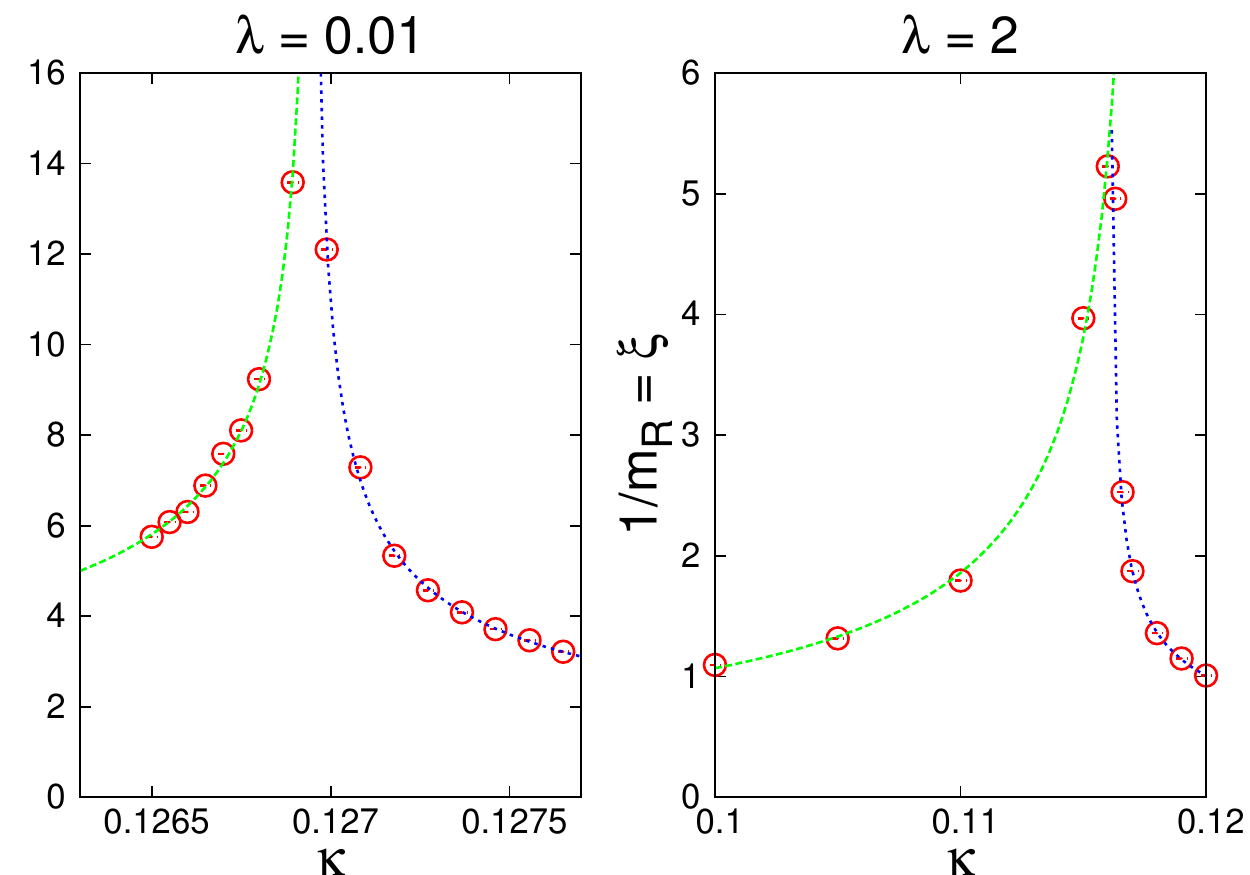}
\caption{Results for $d=4$, $\lambda=0.01$ (\emph{left panel}) and $\lambda=2$ (\emph{right panel}), showing the correlation length, $\xi=1/m_R$, obtained from DMFT as a function of the coupling $\kappa$. The dotted lines are the best fits with exponents $\nu = 0.41$ $(0.47)$ for $\lambda = 0.01$ and $\nu = 0.65$ $(0.43)$ for $\lambda = 2$ in the symmetric (broken-symmetry) phase, respectively.
}
  \label{fig:xi_4d}
\end{figure}

We see that DMFT works well in both cases, although as $\lambda$ increases it deviates more from the exact, mean-field values ($\beta=\nu=1/2$). This can be understood as a consequence of neglecting higher order correlators of the external fields in the expansion of $\exp(-\Delta S)$, Eq.~\eqref{eq:delta_S}, which become more important as the quartic coupling increases. The agreement between EMFT and DMFT is also very good. We see clearly a first order transition in the EMFT result for the stronger coupling, which can explain the deviation of $\beta$ from the mean field value. We do not explicitly see a first order transition in the DMFT result but the convergence of the self-consistency equations for small $\phi_\text{ext}$ is quite poor so we cannot rule it out. Also, based on the good agreement of EMFT and DMFT in three dimensions where both methods predict a first order transition, we suspect that this is also true in four dimensions. Another source of deviation from the mean field exponent might be logarithmic corrections. If we study the divergence of the correlation length for the same couplings we find the behavior shown in Fig.~\ref{fig:xi_4d}. It should be noted that we obtain slightly different values of $\kappa_c$ depending on the phase in which we fit the scaling  behavior of the correlation length.
We find for $\lambda=2$, in the symmetric phase $\kappa_c=0.1175(3)$, $\nu = 0.65(2)$, and in the broken-symmetry phase $\kappa_c=0.1161(2)$, $\nu=0.426(2)$. For $\lambda=0.01$, in the symmetric phase $\kappa_c=0.1269470(5)$, $\nu = 0.41(1)$, and in the broken-symmetry phase $\kappa_c=0.1269475(5)$, $\nu=0.47(1)$.
The results of the fits are summarized in Table~\ref{tab:exponents_4d}. 

\begin{table}
\begin{center}
\caption{\label{tab:exponents_4d}Comparison of critical exponents and critical couplings obtained with Mean Field theory, DMFT, EMFT and Monte Carlo simulations for $\varphi^4$-theory in $d=4$.
The exact values of the critical exponents are the mean-field values, since
$d=4$ is the upper critical dimension.
}
\begin{tabular}{ l | c | c | c }
{} & $\kappa_c$ & $\beta$ & $\nu$\\\hline
\multicolumn{4}{c}{$\lambda = 0.01$}\\\hline
MF &  0.126149 & 1/2 & 1/2 \\
DMFT & 0.12695(1) & 0.508(5)  & 0.47(1) \\
EMFT & 0.1269552(1) & 0.49(2)  & 0.49(3) \\
MC &  0.126945(5) & 0.50(3) & 0.64(4) \\\hline
\multicolumn{4}{c}{$\lambda = 0.1$}\\\hline
MF &  0.131651 & 1/2 & 1/2 \\
DMFT &  0.13655(5) & 0.431(3) & 0.404(3) \\
EMFT &  0.13658(2) & 0.44(4) & 0.45(4) \\
MC &  0.13637(2) & 0.46(7) & 0.57(5) \\\hline
\multicolumn{4}{c}{$\lambda = 0.5$}\\\hline
MF &  0.130756 & 1/2 & 1/2 \\
DMFT &  0.14251(5) & 0.350(2) & 0.51(2) \\
EMFT &  0.14243(2) & 0.36(2) & 0.51(3) \\
MC &  0.1415(1) & 0.45(6)  & 0.51(6) \\\hline
\multicolumn{4}{c}{$\lambda = 2.0$}\\\hline
MF &  0.100313 & 1/2 & 1/2 \\
DMFT &  0.1163(3) & 0.325(4) & 0.426(2) \\
EMFT &  0.11670(5) & 0.29(1) & 0.64(5) \\
MC &  0.1144(5) & 0.44(7)  & 0.50(3) \\\hline
\end{tabular}
\end{center}
\end{table}

\begin{figure}
\includegraphics[width=1\linewidth]{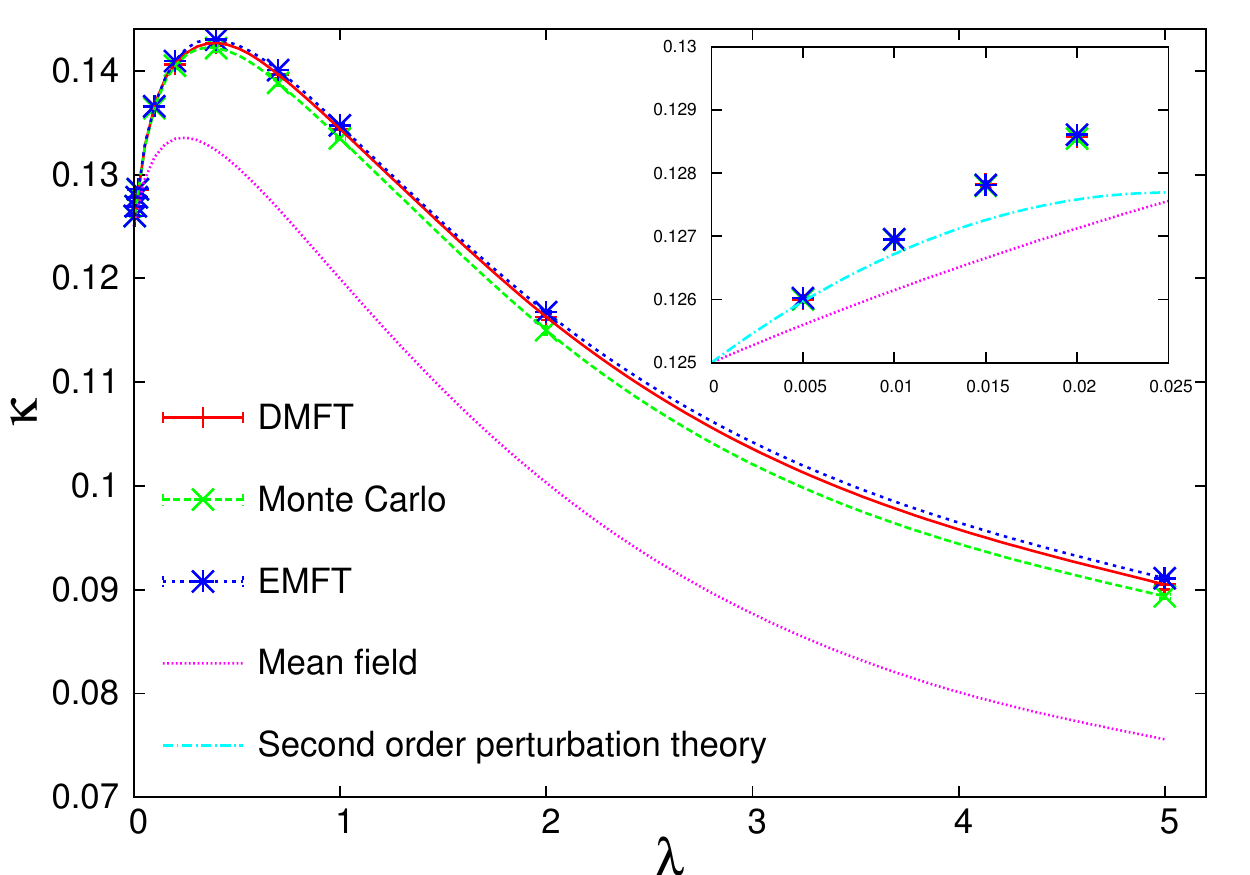}
\caption{Phase diagram of $\varphi^4$-theory 
in four dimensions in the space of $\kappa$ and $\lambda$ obtained by DMFT, EMFT, Monte Carlo and mean field theory. The inset shows the $\lambda \ll 1$ regime where second order perturbation theory also can be used.}
  \label{fig:phi4_phasediag_complete}
\end{figure}

In Fig.~\ref{fig:phi4_phasediag_complete} we show the phase diagram in the $(\kappa,\lambda)$-plane obtained from DMFT, EMFT, Monte Carlo simulation of the full theory, and mean field theory. For $\lambda\ll1$ we have also included results from second order perturbation theory (see inset). In all cases, DMFT is the superior approximation with EMFT close behind. Over the whole range $0\le \lambda\le 5$ it predicts the phase boundary with an accuracy of about 1$\%$ although the transition for larger $\lambda$ is weakly first order.

\begin{figure}[h!]
\centering
\includegraphics[width=1\linewidth]{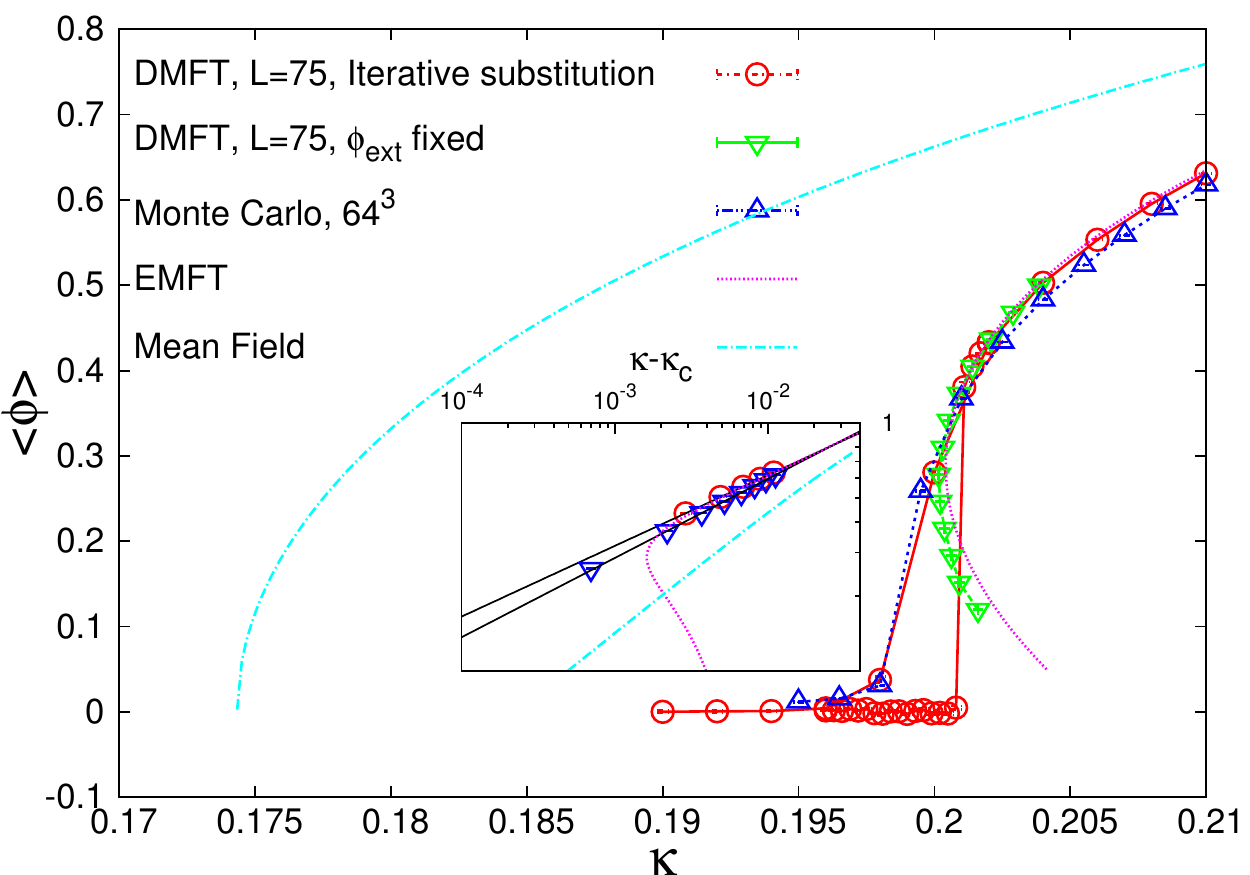}
\caption{Results for $d=3$, $\lambda=0.5$. Expectation value of $\varphi$ as a function of the coupling, $\kappa$, obtained by DMFT, Monte Carlo and mean field theory. The inset shows a log-log plot in the broken-symmetry phase. The slope of the line gives the critical exponent $\beta$.  
}
  \label{fig:expv_3d_l05}
\end{figure}

\noindent
$\bullet$ In two and three dimensions we can explicitly see a first order transition in the DMFT results. This can be clearly seen in Fig.~\ref{fig:expv_3d_l05} which shows the coexistence region (for $d=3$ and $\lambda=1$). The hysteresis in the curve (red, with circles), obtained by iterative substitution of $\phi_\text{ext}$, was obtained by successively increasing or decreasing $\kappa$, using the previous converged $\langle \varphi\rangle$ as input at the next value of $\kappa$ cf.~Fig~\ref{fig:dmft_flow}. For the curve obtained by fixing $\phi_{\text{ext}}$, where a root solver is applied to find the self-consistent $\kappa$, we see that we agree with the substitution method for the stable non-zero solution but that we are also able to obtain the unstable solution. In two dimensions we find a similar situation with an even larger coexistence region.

In order to determine which solution is the physical
solution we have to compare the free energies of the lattice model. 
In DMFT, the free energy of the lattice model can be expressed as the
free energy of the impurity model plus some correction terms \cite{Georges:lec_notes_dmft}. 
The free energy difference in the impurity model, $\delta f$, can be obtained
by taking the logarithm of the ratio of the partition functions,
\bea
\delta f &=& -\log\frac{Z_1}{Z_2} = -\log \frac{\int\mathcal{D}[\varphi]\exp(-S_1)}{\int\mathcal{D}[\varphi]\exp(-S_2)} \nonumber\\
&=& -\log\frac{\int\mathcal{D}[\varphi]\exp(-S_1+S_2)\exp(-S_2)}{\int\mathcal{D}[\varphi]\exp(-S_2)} \nonumber\\
&\equiv& -\log\expv{\exp(-S_1+S_2)}_{Z_2},
\eea
or introducing additional partition functions interpolating
between $Z_1$ and $Z_2$ if necessary \cite{deForcrand:2000fi}. Thus, we can
obtain $\delta f$ by sampling the exponential of the difference
in actions with respect to one of the actions. It turns
out that the solution with $\expv{\varphi}\neq0$ has the lower
impurity free energy. We have not explicitly evaluated the correction 
terms for the DMFT lattice free energies. 

In the EMFT set-up, which gives almost identical results, the correction terms 
have been explicitly worked out in Ref.~\cite{Pankov:thesis}. It turns out that the 
non-zero solution starts as a local minimum for small $\kappa$ and becomes a global minimum when
$\kappa$  is further increased. This suggests that the first few broken-symmetry points in the 
re-entrance region of the DMFT curve may be unstable and justifies disregarding
the data points on the magnetization curve close to where $\partial \kappa/\partial \expv{\varphi} = 0$: it makes sense to use the non-zero branch to estimate $\kappa_c=0.1991(2)$ and $\beta=0.285(1)$ by extrapolating the expectation value. This should be compared with $\kappa_c=0.1988(3)$ and $\beta=0.3200(20)$ from the Monte Carlo data and  $\kappa_c=0.174342(1)$ and $\beta=1/2$ from mean field theory. The value of $\beta$ from the literature is $\beta=0.3267(10)$ \cite{Blote:1999zu}. In Table~\ref{tab:exponents_3d} we summarize the measured quantities for two and three dimensions. Although the strength of the first order transition increases as $\lambda$ increases, the values for $\kappa_c$ and $\beta$ obtained by extrapolating from the broken-symmetry phase are of a comparable quality for the complete range of $\lambda$. In the Monte Carlo data we see that $\beta$ is too large for small $\lambda$, but approaches the correct value with increasing $\lambda$. This is because of the Ginzburg-criterion which states that we will see a mean field-like behavior as soon as the fluctuations of the order parameter are much smaller than the order parameter itself. That means that we have to go very close to the phase transition to see the correct exponents and $\expv{\varphi}$ increases more rapidly when $\lambda$ is small. Because of the strong first order transition we are unable to measure $\nu$ in two and three dimensions.
\begin{table}
\begin{center}
\caption{\label{tab:exponents_3d}Comparison of critical exponents and critical couplings obtained with Mean Field theory, DMFT, EMFT and Monte Carlo simulations for the two and three dimensional $\varphi^4$-theory.
The exact value of $\beta$ (obtained by Monte Carlo in $d=3$) is 0.3267(10)\cite{Blote:1999zu} in $d=3$ and
1/8 in $d=2$.
}
\begin{tabular}{ l | c | c | c | c | c }
{} & \multicolumn{2}{c|}{d=2} &\hspace{0.2cm}& \multicolumn{2}{c}{d=3}\\\hline
{} & $\kappa_c$ & $\beta$ &\hspace{0.2cm} & $\kappa_c$ & $\beta$\\\hline
\multicolumn{6}{c}{$\lambda = 0.01$}\\\hline
MF &  0.252297 & 1/2 &\hspace{0.2cm}& 0.168198 & 1/2\\
DMFT & 0.2610(5) & 0.23(3) &\hspace{0.2cm}&  0.1704(1) & 0.37(4) \\
EMFT & 0.2602(1) & 0.16(3) &\hspace{0.2cm}&  0.1704(1) & 0.36(3) \\
MC &  0.2618(2) & 0.13(1) &\hspace{0.2cm}& 0.17026(3) & 0.44(2) \\\hline
\multicolumn{6}{c}{$\lambda = 0.1$}\\\hline
MF &  0.263301 & 1/2 &\hspace{0.1cm}& 0.175534 & 1/2\\
DMFT &  0.2986(3) & 0.15(2) &\hspace{0.2cm}& 0.1871(1) & 0.32(4) \\
EMFT &  0.2967(5) & 0.15(3) &\hspace{0.2cm}& 0.1872(4) & 0.30(4) \\
MC &  0.3033 & 0.1290(5) &\hspace{0.2cm}& 0.18670(5) & 0.353(3) \\\hline
\multicolumn{6}{c}{$\lambda = 0.5$}\\\hline
MF &  0.261512 & 1/2 &\hspace{0.2cm}& 0.174342 & 1/2\\
DMFT &  0.3305(4) & 0.170(5) &\hspace{0.1cm}& 0.1991(2) & 0.285(10) \\
EMFT &  0.3277(4) & 0.145(20) &\hspace{0.1cm}& 0.1992(3) & 0.28(3) \\
MC &  0.3438(2) & 0.127(3) &\hspace{0.1cm}& 0.1988(3) & 0.3200(20) \\\hline
\multicolumn{6}{c}{$\lambda = 1.0$}\\\hline
MF &  0.23997 & 1/2 &\hspace{0.2cm}& 0.15998 & 1/2\\
DMFT &  0.322(2) & 0.163(9) &\hspace{0.2cm}& 0.1909(3) & 0.23(2) \\
EMFT &  0.318(1) & 0.140(10) &\hspace{0.2cm}& 0.1910(5) & 0.23(2) \\
MC &  0.3402(4) & 0.125(5) &\hspace{0.2cm}& 0.18993(4) & 0.320(4) \\\hline
\end{tabular}
\end{center}
\end{table}

\subsection{Comparison to a cluster variation method}
To quantify the improvement over standard mean field calculations, we compare our results for the critical coupling with Kikuchi's method \cite{Kikuchi:1951}. Kikuchi's method involves three sub-clusters, a single site, a pair of nearest neighbors and a plaquette, as well as three self-consistently determined mean fields. For definiteness we choose a rather large value of $\lambda=5$ and four dimensions. In this region the deviation from the Monte Carlo result is large and the large quartic coupling also helps to keep the four-site integral in Kikuchi's method numerically manageable. The results can be found in Table~\ref{tab:comp_kikuchi}.

\begin{table}[t]
\begin{center}
\caption{\label{tab:comp_kikuchi}Illustrative comparison of critical couplings obtained with Mean Field theory (MF), DMFT, EMFT, Kikuchi's method and Monte Carlo simulations for the four dimensional $\varphi^4$-theory at $\lambda=5$.
}
\begin{tabular}{ c | c | c | c | c }
MF & DMFT & EMFT & Kikuchi & MC\\\hline
$0.07521$ & $0.09007(10)$ & $0.09064(20)$ & $0.08859$ & $0.08893(20)$\\
\end{tabular}
\end{center}
\end{table}

While Kikuchi's method gives a slightly better estimate for the critical coupling,  the accuracy of the DMFT prediction is comparable. At smaller values of $\lambda$ it becomes increasingly difficult to keep the numerical errors in Kikuchi's method under control, whereas DMFT and EMFT do not suffer from any convergence problem. 

\section{Summary and Outlook}\label{sec:sum}\noindent
To summarize, we found that the DMFT approximation in some aspects provides a remarkably accurate description of $\varphi^4$-theory in dimensions $d\geq2$, especially for small quartic coupling and high dimensions. A posteriori, this is quite natural:
DMFT is rooted in a mean-field approach, which works better in high dimension,
and in an approximation for the self-energy, which works better if
interactions are weak. Within these limitations, DMFT is a remarkable
improvement over ordinary mean-field theory: with modest computer resources, it
provides an estimate of the critical coupling $\kappa_c(\lambda)$ to an
accuracy ${\mathcal O}(10^{-5})$ (see Fig.~6), and reasonably accurate critical 
exponents. In addition, DMFT yields an approximation of the Euclidean
two-point function, from which one may extract the spectral density or
the real-time, analytically continued correlator.
The finite-temperature behavior can conveniently be studied as well, by
varying the extent of the preserved compact, dynamical dimension.
Such information is out of reach of the standard mean-field approach.

DMFT incorrectly predicts a first-order transition for all 
values of the quartic coupling in dimension $d\leq3$. In four dimensions 
we find a second order transition only for very small quartic coupling.
This kind of breakdown below the upper critical dimension is typical for
mean field like methods.
It is only when $d\geq5$ that the transition is of second order for all 
couplings. A first-order transition means that we cannot take the continuum 
limit of the lattice theory. Nevertheless, an effective scaling behavior, with
associated effective critical exponents, can be observed in the regime of 
large correlation length near the weak first-order transition.

We have also shown that, if non-local quantities are not of direct interest,
the local limit of DMFT (extended mean-field theory, EMFT \cite{Pankov:2002}) 
can make predictions which are vastly superior to mean field theory at a low computational cost.

In conclusion, while DMFT might not be reliable in $d=2$ and 3, it is a useful method for the study of $\varphi^4$ theory in $d=4$. Depending on the region in parameter space
and the observables of interest, DMFT offers a remarkably accurate and computationally less expensive alternative to simulations of the full model.

It would be particularly interesting to apply DMFT and EMFT to theories
afflicted by a ``sign problem'', because in that case the computer resources required
for a reliable Monte Carlo study are prohibitive. With this in mind,
the natural next step in our study is the extension to complex $\varphi^4$ theory, 
which has additional interesting features of great physical interest. 
It has a continuous global symmetry which allows us to introduce a chemical potential $\mu$. For sufficiently large $\mu$, the global symmetry is spontaneously broken
and the system undergoes Bose condensation via a phase transition.
This chemical potential leads to a sign problem in Monte Carlo simulations, which has much in common with the sign problem in QCD at finite chemical potential. 
We have derived in Appendix~\ref{app:imp} the impurity action and the self-consistency equations, so if the sign problem in the one-dimensional chain can be dealt with, the simulations should be straightforward. Anders et al. \cite{Anders:2011} have simulated a similar system with a sign problem, so there is hope that DMFT may also give good results for the complex $\varphi^4$-theory, at least in the vicinity of the Bose-condensation phase transition. It is also possible that EMFT will already give good results when applied to complex $\varphi^4$-theory \cite{Akerlund:2013}.

Yet another possible direction to proceed would be to improve the method for the existing model, for instance by pushing the cumulant expansion of $\zeta$, Eq.~\eqref{eq:A8_2}, to higher order. 
This would be a complementary effort to existing work on the systematic improvement of DMFT \cite{Jabben:2012}, 
and cluster extensions of DMFT 
\cite{Maier:2005}.

The most appealing prospect is perhaps cluster DMFT where instead of an impurity model consisting of a single one-dimensional chain one considers a narrow cylinder. At each time point the field is then allowed to fluctuate in the spatial direction, at least on short scales. This would introduce some dependence on short-range fluctuations in space and hence on large spatial momenta in the self-energy, which should improve the approximation. This approach is especially attractive when we think about applications of DMFT to gauge theories. Since the smallest gauge invariant object is a plaquette, it is necessary to treat at least a plaquette (four-site cluster) at each time.

Self-consistency equations involving plaquette variables can be written down, as a generalization of, eg., Ref.~\cite{Drouffe:1983}.
The non-local kernel $\Delta(t-t')$ would now describe plaquette-plaquette interactions. Hopefully, a satisfactory approximation
can be achieved when truncating $\Delta$ to a few terms. The fast decay of $\Delta(t-t')$ observed in the $\varphi^4$ case, even close to the phase transition, fosters optimism.

\acknowledgements
We thank Jens Langelage, Tobias Rindlisbacher, Peter Staar and Wolfgang Unger
for helpful discussions.
A.G. acknowledges the hospitality of the Pauli Center at ETH in the framework of the Schr\"{o}dinger Chair.  
The calculations were performed on the Brutus cluster at ETH Zurich.

\bibliography{bib}

\begin{thebibliography}{10}

\bibitem{Batrouni:1982bg}
G.~G. Batrouni, {\it Nucl.Phys.\/} {\bf B208}, 467 (1982).

\bibitem{Batrouni:1982dx}
G.~G. Batrouni, {\it Nucl.Phys.\/} {\bf B208}, 12 (1982).

\bibitem{Batrouni:1984rc}
G.~G. Batrouni, M.~B.~Halpern, {\it Phys.Rev.\/} {\bf D30}, 1775 (1984).

\bibitem{Kallman:1984ky}
C.~Kallman, {\it Phys.Lett.\/} {\bf B134}, 363 (1984).

\bibitem{Green:1983sd}
F.~Green, F.~Karsch, {\it Nucl.Phys.\/} {\bf B238}, 297 (1984).

\bibitem{Georges:1996zz}
A.~Georges, G.~Kotliar, W.~Krauth, M.~J. Rozenberg, {\it Rev.Mod.Phys.\/} {\bf
  68}, 13 (1996).

\bibitem{Georges:lec_notes_dmft}
A.~Georges, Strongly correlated electron materials: Dynamical mean-field theory
  and electronic structure. Cond-mat/0403123.

\bibitem{Gull:2011}
E.~Gull, {\it et~al.\/}, {\it Rev. Mod. Phys.\/} {\bf 83}, 349 (2011).

\bibitem{Byczuk:2008}
K.~Byczuk, D.~Vollhardt, {\it Phys. Rev. B\/} {\bf 77}, 235106 (2008).

\bibitem{Anders:2010}
P.~Anders, E.~Gull, L.~Pollet, M.~Troyer, P.~Werner, {\it Phys. Rev. Lett.\/}
  {\bf 105}, 096402 (2010).

\bibitem{Anders:2011}
P.~Anders, E.~Gull, L.~Pollet, M.~Troyer, P.~Werner, {\it New Journal of
  Physics\/} {\bf 13}, 075013 (2011).

\bibitem{Anders:2012}
P.~Anders, P.~Werner, M.~Troyer, M.~Sigrist, L.~Pollet, {\it Phys. Rev.
  Lett.\/} {\bf 109}, 206401 (2012).

\bibitem{Pankov:2002}
S.~Pankov, G.~Kotliar, Y.~Motome, {\it Phys. Rev. B\/} {\bf 66}, 045117 (2002).

\bibitem{Bethe:1935}
H.~A.~Bethe, {\it Proc. Roy. Soc. London A} {\bf 150} 552, (1935).

\bibitem{Kikuchi:1951}
R.~Kikuchi, {\it Phys. Rev.} {\bf 81}, 988 (1951).

\bibitem{Morita:1994}
T.~Morita, {\it Prog. Theor. Phys.} {\bf 115}, 27 (1994).



\bibitem{Matsuo:2006}
T.~Matsuo, Y.~Natsume, T.~Kato, {\it J. Phys. Soc. Jpn.\/} {\bf 75}, 103002
  (2006).

\bibitem{Werner:2005}
P.~Werner, M.~Troyer, {\it Phys. Rev. Lett.\/} {\bf 95}, 060201 (2005).

\bibitem{Strand:2011}
H.~U.~R. Strand, A.~Sabashvili, M.~Granath, B.~Hellsing, S.~\"Ostlund, {\it
  Phys. Rev. B\/} {\bf 83}, 205136 (2011).

\bibitem{Dyson:1969up}
F.~Dyson, {\it Commun.Math.Phys.\/} {\bf 12}, 91 (1969).

\bibitem{Dyson:1969up_2}
F.~Dyson, {\it Commun.Math.Phys.\/} {\bf 12}, 212 (1969).

\bibitem{Frohlich:1982gf}
J.~Fr\"ohlich, T.~Spencer, {\it Commun.Math.Phys.\/} {\bf 83}, 411 (1982).

\bibitem{Fisher:1972zz}
M.~E. Fisher, S.-k. Ma, B.~Nickel, {\it Phys.Rev.Lett.\/} {\bf 29}, 917 (1972).

\bibitem{Luijten:2001}
E.~Luijten, H.~Me\ss{}ingfeld, {\it Phys. Rev. Lett.\/} {\bf 86}, 5305 (2001).

\bibitem{Wolff:1989}
U.~Wolff, {\it Phys. Lett. B\/} {\bf 228}, 379  (1989).

\bibitem{deForcrand:2000fi}
P.~de~Forcrand, M.~D'Elia, M.~Pepe, {\it Phys.Rev.Lett.\/} {\bf 86}, 1438
  (2001).

\bibitem{Pankov:thesis}
S.~Pankov, Dynamical mean field theories and the anderson localization, Ph.D.
  thesis, Rutgers, The State University of New Jersey (2003).

\bibitem{Blote:1999zu}
H.~W.~J.~Bl\"ote, J.~R.~Heringa, M.~M.~Tsypin, {\it Phys.Rev.\/} {\bf E62}, 77 (2000).

\bibitem{Akerlund:2013}
O.~Akerlund, P.~de~Forcrand, in proceedings of ``31st International Symposium on Lattice Field Theory'', PoS(LATT2013)195.

\bibitem{Jabben:2012}
T.~Jabben, N.~Grewe, S.~Schmitt, {\it Phys. Rev. B\/} {\bf 85}, 165122 (2012).

\bibitem{Maier:2005}
T.~Maier, M.~Jarrell, T.~Pruschke, M.~H. Hettler, {\it Rev. Mod. Phys.\/} {\bf
  77}, 1027 (2005).

\bibitem{Kotliar:2006}
G.~Kotliar, {\it et~al.\/}, {\it Rev. Mod. Phys.\/} {\bf 78}, 865 (2006).

\bibitem{Drouffe:1983}
J.~M.~Drouffe, J.~B.~Zuber, {\it Phys. Rep.\/} {\bf 102}, 1 (1983).

\end{thebibliography}
\bibliographystyle{Science}

\appendix
\section{Impurity action of complex $\varphi^4$-theory}\label{app:imp}\noindent
In this appendix, we derive the DMFT equations for the complex valued, scalar $\varphi^4$-theory in the presence of a chemical potential $\mu$. This is more general than the real-valued, scalar $\varphi^4$ theory, studied in this paper, which can be obtained as a special case of the complex theory. We derive the action and self-consistency equations using an effective medium approach, closely following Ref.~\cite{Anders:2011}. We assume that the global symmetry, ($U(1)$ when the field is complex and $\mathbb{Z}_2$ in the real case),  is broken since this is the more general case, and since the equations for the symmetric phase can be easily obtained by setting the expectation value of the field to zero. The expectation value of the field in the broken-symmetry phase will then be determined self-consistently as in standard Mean Field theory.

We start by quickly reminding the reader of the lattice action of complex $\varphi^4$-theory with a chemical potential $\mu$ coupled to the time component of the conserved current:
\begin{align}\label{eq:app_action_latt}
S &= \sum_x\left(\abs{\varphi_x}^2 + \lambda(\abs{\varphi_x}^2-1) \phantom{\sum_\nu^4}\right.\\
&\phantom{=}\left.-\kappa\sum_\nu\left[e^{-\mu\delta_{\nu,t}}\varphi^*_x\varphi_{x+\hat{\nu}} + e^{\mu\delta_{\nu,t}}\varphi^*_x\varphi_{x-\hat{\nu}}\right]\right),\nonumber
\end{align}
where $\delta_{\nu,t}=1$ if $\nu$ is the time direction and zero otherwise.

We consider a lattice with $N_s^{d-1}N_t$ sites ($N_s$ can formally be taken to be infinite) and denote by $\varphi_{\vec{i},t}$ the field on the site $(\vec{i},t) = (x_1,\ldots,x_{d-1},t)$. We then single out $\vec{i}=\vec{0}\equiv (0,\ldots,0)$ and call the world line at the spatial origin, $\varphi_{\text{int},t}\equiv\varphi_{\vec{0},t}$, the \emph{internal} degrees of freedom. All other sites are considered an external effective bath or \emph{external} degrees of freedom, $\varphi_{\text{ext},t} = \{\varphi_{\vec{j},t} : \vec{j}\neq \vec{0}\}$. We will also use the Nambu notation throughout. In the Nambu notation we have $\bphi^\dagger = (\varphi^*,\varphi)$ and this allows us to write equations with $\varphi$ and $\varphi^*$ more compactly using vectors and matrices.

We can write the action, Eq.~\eqref{eq:app_action_latt}, as a sum of three terms, the action of the world line, the action of the external sites and the interaction of the world line with the external bath. The action in the Nambu notation reads
\begin{align}
S &= \displaystyle\sum_x\left[\!-\kappa\sum_\nu \bphi^\dagger_{x +\widehat{\nu}}\bm{E}(\mu\delta_{\nu,t})\bphi_x \!+\! \frac{1}{2}\abs{\bphi_x}^2 \!+\! \frac{\lambda}{4}(\abs{\bphi_x}^2\!-\!2)^2\!\right] \nonumber\\
&= S_\text{int} + \Delta S + S_\text{ext},
\end{align}
with 
\begin{widetext}
\begin{align}
\bm{E}(x) &= \begin{pmatrix} e^{-x} & 0 \\ 0 & e^{x} \end{pmatrix}, \\
S_\text{int} &=  \displaystyle\sum_t\left[-\kappa\bphi^\dagger_{\text{int},t +1}\bm{E}(\mu)\bphi_{\text{int},t} + \frac{1}{2}\abs{\bphi_{\text{int},t}}^2 + \frac{\lambda}{4}(\abs{\bphi_{\text{int},t}}^2-2)^2\right],\\
\Delta S &= -\kappa \displaystyle\sum_t \sum_{\langle \text{int},\text{ext}\rangle} \bphi^\dagger_{\text{int},t}\bphi_{\text{ext},t}, \\
S_\text{ext} &=  \displaystyle\sum_{x\neq(\vec{0},t)}\left[-\kappa\sum_{\substack{\nu\\x+\hat\nu\neq(\vec{0},t)}} \bphi^\dagger_{x +\widehat{\nu}}\bm{E}(\mu\delta_{\nu,t})\bphi_x + \frac{1}{2}\abs{\bphi_x}^2 + \frac{\lambda}{4}(\abs{\bphi_x}^2-2)^2\right].
\end{align}
\end{widetext}

The sum over $\langle\text{int,ext}\rangle$ is shorthand for the sum over all external sites at time $t$ which are nearest neighbors to the internal site at time $t$.

The surrounding bath is considered to be of infinite size and can thus spontaneously break the symmetry and develop an expectation value. The world-line subject to the action $S_\text{int}$ can not spontaneously break the symmetry, since $d=1$ is the lower critical dimension of the Ising universality class, (and $d=2$ for a $U(1)$ symmetry). At and below the lower critical dimension, the system is always disordered since the entropy gain of introducing a domain wall wins over the energy cost. In a one-dimensional chain of Ising spins the cost of breaking one bond is constant but it does not cost any energy to move the broken bond along the chain. However, one-dimensional systems can exhibit a phase transition if there are long-range interactions \cite{Dyson:1969up, Dyson:1969up_2, Frohlich:1982gf, Fisher:1972zz, Luijten:2001}. Such long-range interactions can be induced as the field on the chain interacts with the bath through $\Delta S$ and also $\bphi_\text{int}$ can acquire a non-zero expectation value. In order to account for this possibility we write, 
\beann
\bphi_{\text{ext},t} &=& \bm{\phi}_\text{ext} + \bdphi_{\text{ext},t}, \,\,\, \langle \bphi_{\text{ext},t}\rangle = \bm{\phi}_\text{ext},\\
\bphi_{\text{int},t} &=& \bm{\phi}_\text{int} + \bdphi_{\text{int},t}, \,\,\, \langle \bphi_{\text{int},t}\rangle = \bm{\phi}_\text{int}.
\eeann
Notice that the two expectation values are not dynamical variables but rather constants that can be tuned to achieve self-consistency. Inserting this in $ \Delta S $ yields,
\begin{align}
\Delta S =& -\kappa \displaystyle\sum_t\Bigg(2(d-1)\bm{\phi}^\dagger_\text{ext}\bdphi_{\text{int},t} \nonumber\\
&\hspace{5mm}+ \bdphi^\dagger_{\text{int},t}\sum_{\langle \text{int},\text{ext}\rangle}\bdphi_{\text{ext},t} + 2(d-1)\bm{\phi}^\dagger_\text{int}\bdphi_{\text{ext},t}\Bigg).   
\end{align}
There are three different terms which are dealt with differently. The first term can be included in $S_\text{int}$. We assume small fluctuations around the classical solution so the second term can be used to expand the Boltzmann weight. The third term is independent of the internal degrees of freedom ($\phi_\text{int}$ is considered fixed) and is included in $S_\text{ext}$. Let us define
\begin{align}
S_1 &= -2\kappa (d-1)\displaystyle\sum_t\bm{\phi}^{\dagger}_\text{ext}\bdphi_{\text{int},t}, \\
\delta S &= -\kappa \displaystyle\sum_t \sum_{\langle \text{int},\text{ext}\rangle} \bdphi^\dagger_{\text{int},t}\bdphi_{\text{ext},t} \equiv \sum_t\delta S(t),
\end{align}
and expand $ \exp(-\delta S) $ to get
\begin{align}
Z &= Z_\text{ext}\int\mathcal{D}\varphi_\text{int}\exp(-S_\text{int} - S_1)\zeta,\nonumber\\
\zeta &= 1 - \sum_t \langle\delta S(t) \rangle_\text{ext} + \frac{1}{2}\sum_{t,t'}\langle\delta S(t)\delta S(t')  \rangle_\text{ext} + \ldots,\label{eq:A8_2}
\end{align}
where $Z_\text{ext}\equiv\int\mathcal{D}\exp(-S_\text{ext})$ is the partition function of the action including only $\varphi_{\text{ext},t}$. The expectation values are with respect to $Z_\text{ext}$. The first order term in $\zeta$ is proportional to the expectation value of $\delta\varphi_\text{ext}$ which is zero by construction. The second-order term is non-zero and we find,
\begin{align}
&\langle\delta S(t)\delta S(t')  \rangle_\text{ext}  \nonumber\\
&\hspace{5mm}=\kappa^2\bdphi^\dagger_{\text{int},t}\!\sum_{\langle \text{int},\text{ext}\rangle}\sideset{}{'}\sum_{\langle \text{int},\text{ext}\rangle}\expv{\bdphi_{\text{ext},t}\bdphi^\dagger_{\text{ext},t'}}_\text{ext}\bdphi_{\text{int},t'},\nonumber\\
&\hspace{5mm} = \bdphi^\dagger_{\text{int},t}\bm{\Delta}(t-t')\bdphi_{\text{int},t'}
\end{align}
where the prime on the second sum means that there are two different external sites.
This corresponds to a field propagating in the effective medium subjected to the unknown propagator, between creation and annihilation at the spatial origin. This term is called ``hybridization function" $\bm{\Delta}(t-t')$. It originates from connected diagrams and will be determined self-consistently in the DMFT-loop by demanding that the local Green's function of the effective model coincide with the local Green's function of the full model. $\bm{\Delta}(t)$ is a $2\times2$ hermitian matrix and there is of course an implicit dependence on $\mu$. Here we can see that DMFT is a better approximation at high dimensionality. We have already argued that mean field theory should be exact in $d=\infty$ and in mean field theory quadratic fluctuations are completely ignored. That means that expectation values like $\langle\bdphi_{\text{ext},t}\bdphi^\dagger_{\text{ext},t'}\rangle_\text{ext}$ factorize and there is no error involved when we neglect contributions from higher order correlators. The diagonal entry of $\bm{\Delta}(t)$ is associated with $\langle\varphi^*(0)\varphi(t)\rangle$ and the off-diagonal entry is associated with $\langle\varphi(0)\varphi(t)\rangle$. After the re-exponentiation Eq.~\eqref{eq:A8_2} we find the impurity action,
\begin{widetext}
\begin{align}
S_\text{imp} &= \displaystyle\sum_t\left[-\kappa\bphi^\dagger_{t +1}\bm{E}(\mu)\bphi_t + \frac{1}{2}\abs{\bphi_t}^2 + \frac{\lambda}{4}(\abs{\bphi_t}^2-2)^2\right] 
- \frac{1}{2}\sum_{t,t'} \bdphi^\dagger_t \bm{\Delta}(t-t')\bdphi_{t'} - 2\kappa(d-1)\bm{\phi}^\dagger\sum_t\bphi_t\nonumber\\
&= \displaystyle\sum_t\left[-\kappa\bphi^\dagger_{t +1}\bm{E}(\mu)\bphi_t + \frac{1}{2}\abs{\bphi_t}^2 + \frac{\lambda}{4}(\abs{\bphi_t}^2-2)^2\right] 
-\frac{1}{2} \sum_{t,t'} \bphi^\dagger_t \bm{\Delta}(t-t')\bphi_{t'} - \bm{\phi}^\dagger\left(2\kappa(d-1)\bm{I}-\widetilde{\bm{\Delta}}(0)\right)\sum_t\bphi_t\nonumber\\
&= \sum_{t,t'}\bphi^\dagger_t {\bm{K}}_{\text{imp},c}^{-1}(t-t')\bphi_{t'} + \lambda\displaystyle\sum_t (\abs{\varphi_{t}}^2-1)^2 - \bm{h}^\dagger\sum_t\bphi_t,
\end{align}
\end{widetext}
where $\widetilde{\bm{K}}_{\text{imp},c}^{-1}(\omega) = \tfrac{1}{2}\bm{I}-\kappa\bm{E}(\mu-i\omega) - \tfrac{1}{2}\widetilde{\bm{\Delta}}(\omega)$ is the inverse of the connected two-point Green's function of the free theory and $\bm{h}^\dagger = \bm\phi^\dagger\left(2\kappa(d-1)\bm{I}-\widetilde{\bm{\Delta}}(0)\right)$ plays the role of an external magnetic field. Note that the factor of $1/4$ that sometimes appears in front of $\lambda$ is due to the identity $\abs{\bphi}^2 = 2\abs{\varphi}^2$. Setting $\bphi$ to be real and the chemical potential to be zero, one recovers Eqs.~(\ref{eq:simp_1}-\ref{eq:simp_3}).

\section{Extended Mean Field Theory}\label{app:sdmft}
\noindent
In the zero-dimensional, or local, model we have the impurity action
\be
S_\text{imp}= (1-\Delta)\varphi^2 -2\phi_\text{ext}(2d\kappa-\Delta)\varphi + \lambda(\varphi^2-1)^2.
\ee
Since we are working with a single site model in coordinate space all quantities are local in space and time, and all quantities of the full lattice model which enter the self-consistency equations are understood to be local in space and time as well. Close to the phase transition, where $\phi_\text{ext}$ is very small, we can expand the exponential of the action in powers of $\phi_\text{ext}$. It is convenient to define,
\begin{align}
Z_0 &\equiv \int \rd \varphi \exp\left(-S_\text{imp}|_{\phi_\text{ext}=0}\right),\\
M_k &\equiv \expv{\varphi^k}_{Z_0}.
\end{align}
In $Z_0$ we only discard the explicit dependence on $\phi_\text{ext}$ but not the implicit dependence in $\Delta$. In this setup $M_k$ actually depends on $\phi_\text{ext}$. A naive expansion to order $\mathcal{O}\left(\phi_\text{ext}^4\right)$ gives:
\begin{align}
\langle\varphi\rangle &= 2(2d\kappa-\Delta)\phi_\text{ext}M_2 \nonumber\\ &\phantom{=}+\frac{8}{6}(2d\kappa-\Delta)^3\phi_\text{ext}^3\left(M_4-3M_2^2\right),\\
G_\text{imp} &= 2\left(M_2+2\phi_\text{ext}^2(2d\kappa-\Delta)^2\left(M_4-3M_2^2\right)\right).
\end{align}
Using the self-consistency condition $\langle\varphi\rangle=\phi_\text{ext}$ we can determine $\Delta$ up to order $\phi_\text{ext}^2$,
\be
\Delta = 2d\kappa - \frac{1}{2M_2} + \frac{\phi_\text{ext}^2}{12M_2^2}\left(\frac{M_4}{M_2^2}-3\right),
\ee
which leads to an effective action in terms of $\kappa,\phi_\text{ext}$ and $M_2,M_4$, which needs to be determined self-consistently. More precisely,
\begin{align}
S_\text{imp} &= \left(1-2d\kappa+\frac{1}{2M_2} - \frac{\phi_\text{ext}^2}{12M_2^2}\left(\frac{M_4}{M_2^2}-3\right)\right)\varphi^2 \nonumber\\&\phantom{=}+ \lambda\left(\varphi^2-1\right)^2,\label{eq:simp_exp}\\
G_\text{loc} &= \int_0^\infty\rd \tau \exp\left[-\tau\left(2d\kappa+\frac{\phi_\text{ext}^2}{6M_2^2}\left(3-\frac{M_4}{M_2^2}\right)\right)\right]\nonumber\\&\phantom{=}\times I_0(2\kappa\tau)^d,\label{eq:gloc_exp}\\
G_\text{imp} &= 2M_2-\phi_\text{ext}^2\left(3-\frac{M_4}{M_2^2}\right),
\end{align}
which defines our self-consistency equations. $M_4$ already multiplies $\phi_\text{ext}^2$ so, to order $\phi_\text{ext}^2$, we can use its value at $\phi_\text{ext}=0$, i.e. $M_4\equiv M_4|_{\phi_\text{ext} =0}$. However, the $\phi_\text{ext}^2$ correction to $M_2$ contributes and must be calculated. We find,
\be
M_2 = M_2|_{\phi_\text{ext} =0} - \frac{\phi_\text{ext}^2}{6}\left(\frac{M_4}{(M_2|_{\phi_\text{ext} =0})^2}-1\right),
\ee
which gives the Green's function,
\be
G_\text{imp} = 2M_2|_{\phi_\text{ext} =0}-\frac{2\phi_\text{ext}^2}{3}\left(4-\frac{M_4}{(M_2|_{\phi_\text{ext} =0})^2}\right).
\ee
$M_2|_{\phi_\text{ext} =0}$ must be determined self-consistently using the action in Eq.~\eqref{eq:simp_exp}. Equating $G_\text{imp}$ and $G_\text{loc}$ at $\phi_\text{ext} = 0$ yields,
\be
2M_2|_{\phi_\text{ext} =0} = \frac{1}{2\kappa}\int_0^\infty\rd \tau \exp\left(-\tau d\right)I_0(\tau)^d,
\ee
which defines the critical coupling. In the Ising limit the situation simplifies since $M_4=M_2=1$.

A crucial question is how the $\kappa$ which equates $G_\text{imp}$ and $G_\text{loc}$ behaves as a function of $\phi_\text{ext}$. The analysis with arbitrary $\lambda$ is quite involved but we can learn a lot by considering the Ising limit where $\lambda = \infty$. We have to distinguish between two cases: when $d>4$ the integral,
\be
\int_0^\infty\rd \tau\;\tau e^{-\tau d}I_0(\tau)^d \equiv I'_d\label{eq:idp}
\ee
is finite and we can expand the exponential in Eq.~(\ref{eq:gloc_exp}). After some algebra we find,
\begin{align}
\kappa &= \frac{I_d}{4} + \left(\frac{I_d}{4}-\frac{I'_d}{6I_d}\right)\phi_\text{ext}^2,\\
I_d &= \int_0^\infty\rd \tau\; e^{-\tau d}I_0(\tau)^d.
\end{align}
If the coefficient in front of $\phi_\text{ext}^2$ is negative we have a first order transition because we will have a region with multiple solutions and this indeed happens for $d\lsim4.9$. For larger $d$ the coefficient is positive and we find a second order transition with critical exponent $\beta=1/2$.

When $3<d\leq 4$, the integral in Eq.~\eqref{eq:idp} is divergent and we cannot use a Taylor expansion of Eq.~\eqref{eq:gloc_exp}. Instead we can numerically study how Eq.~\eqref{eq:gloc_exp} behaves for small $\phi_\text{ext}$. We restrict ourselves to dimensions larger than three since otherwise $I_d$ also is divergent. We find that $G_\text{loc}$ has an expansion,
\be
G_\text{loc} = \frac{I_d}{2\kappa} - \frac{B}{\kappa^2}\phi_\text{ext}^\alpha - C\phi_\text{ext}^2, \; B,C >0, \; \alpha < 2.
\ee
This means that for small enough $\phi_\text{ext}$ the self-consistency equation takes the form,
\be
\frac{I_d}{2\kappa} - \frac{B}{\kappa_c^2}\phi_\text{ext}^\alpha = 2,
\ee
which gives,
\be
\kappa = \frac{I_d}{4} -32\frac{B}{I_d^3}\phi_\text{ext}^\alpha.
\ee
Again, due to the negative prefactor, the transition is first order. In Fig.~\ref{fig:lambda_tc} we show the quartic coupling for which the transition turns second order as a function of the dimension.

\end{document}